\documentclass[
  aps,prd,reprint,
  amsmath,amssymb,
  superscriptaddress,
  floatfix,nofootinbib
]{revtex4-2}

\usepackage[T1]{fontenc}
\usepackage{newtxtext,newtxmath}
\usepackage{graphicx}
\usepackage{xspace}
\usepackage{orcidlink}
\usepackage[dvipsnames]{xcolor}

\usepackage{hyperref}
\hypersetup{
  colorlinks=true,
  linkcolor=blue,
  citecolor=blue,
  urlcolor=blue
}

\usepackage{aas_macros} 

\usepackage{setspace}

\begin{document}

\newcommand{\hiMpc}{h^{-1} {\rm Mpc}}
\newcommand{\hiMsun}{h^{-1} M_\odot}

\newcommand{\dd}{{\rm d}} 

\newcommand{\redmapper}{redMaPPer\xspace} 
\newcommand{\redmagic}{redMaGiC\xspace}

\newcommand{\zlam}{z_{\lambda}}
\newcommand{\zcl}{z_{\rm cl}}
\newcommand{\zred}{z_{\rm red}}
\newcommand{\zspec}{z_{\rm spec}}
\newcommand{\zr}{z_{\rm r}}

\newcommand{\pmem}{p_{\rm mem}}

\newcommand{\bu}{b_{\rm u}}
\newcommand{\br}{b_{\rm r}}

\newcommand{\phiu}{\phi_{\rm u}}
\newcommand{\phir}{\phi_{\rm r}}

\newcommand{\Du}{{\rm D_{\rm u}}}
\newcommand{\Dr}{{\rm D_{\rm r}}}
\newcommand{\Rr}{{\rm R_{\rm r}}}

\newcommand{\DuDrhat}{\widehat{\Du\Dr}}
\newcommand{\DuRrhat}{\widehat{\Du\Rr}}

\newcommand{\wmm}{w_{\rm mm}}
\newcommand{\wur}{w_{\rm ur}}
\newcommand{\wurhat}{\hat{w}_{\rm ur}}
\newcommand{\phiuhat}{\hat{\phi}_{\rm u}}

\newcommand{\thetamin}{{\theta_{\rm min}}}
\newcommand{\thetamax}{{\theta_{\rm max}}}

\newcommand{\sigmaproj}{\sigma_{\rm proj}}
\newcommand{\sigmacl}{\sigma_{\rm cl}}
\newcommand{\fproj}{f_{\rm proj}}

\newcommand{\sigmav}{\sigma_{\rm v}} 

\title{Calibrating optical galaxy cluster projection effects with sparse spectroscopic samples:\\
A clustering redshift approach}

\author{Lei Yang}
\email{lyang4@smu.edu}
\affiliation{Department of Physics, Southern Methodist University, Dallas, TX 75205, USA}
\author{Hao-Yi Wu}
\email{hywu@smu.edu}
\affiliation{Department of Physics, Southern Methodist University, Dallas, TX 75205, USA}
\author{Tesla Jeltema}
\affiliation{Physics Department, University of California, Santa Cruz, CA 95064, USA\\
Santa Cruz Institute for Particle Physics, Santa Cruz, CA 95064, USA}
\author{Chun-Hao To}
\affiliation{Department of Astronomy and Astrophysics, University of Chicago, Chicago, IL 60637, USA}
\author{Ross Cawthon}
\affiliation{Oxford College of Emory University, Oxford, GA 30054, USA
}
\author{Shulei Cao}
\affiliation{Department of Physics, Southern Methodist University, Dallas, TX 75205, USA}

\date{\today}

\begin{abstract}
Wide-field optical imaging surveys are efficient at identifying galaxy clusters, but optically identified clusters suffer from projection effects---physically unassociated galaxies along the line of sight are often included as cluster members due to distance uncertainties.  Previous studies have used spectroscopic follow-up observations of cluster members to quantify projection effects; however, such follow-up efforts cannot keep pace with the rapidly growing cluster samples.  On the other hand, spectroscopic surveys designed for large-scale structure studies collect tens of millions of spectra but tend to have sparse spectra in cluster regions. To bridge this gap, we develop a clustering redshift approach that cross-correlates cluster members with sparse, noncluster-targeted spectroscopic galaxy samples.  We validate this approach using the Cardinal simulation, recovering the correct spectroscopic redshift distribution and projection effect parameters of cluster members identified by the red-sequence Matched-filter Probabilistic Percolation algorithm.  Our approach is insensitive to the selection of the spectroscopic sample and paves the way for calibrating projection effects in the upcoming Legacy Survey of Space and Time clusters using DESI and Roman spectroscopic samples.
\end{abstract}
\maketitle

\section{Introduction}

The abundances of galaxy clusters across cosmic time are sensitive to both the cosmic expansion history and the growth of structure \cite{Haiman01, Holder01, LimaHu04, Voit05, Allen11, Weinberg13, Huterer15}.  To measure cluster abundances, cluster surveys have been conducted in optical \cite{Gladders07, Koester07b, Rykoff14, Rykoff16, Oguri18HSC-CAMIRA, Maturi25KiDS, DESY3CL3x2pt}, x-ray \cite{Bohringer04, Liu22eFEDS, Bulbul24}, and millimeter \cite{Planck15Cluster, Bleem24, ACT26} bands.  Among different survey techniques, wide-field optical imaging surveys, including the Vera C.~Rubin Observatory's Legacy Survey of Space and Time (LSST) \cite{Ivezic19}, Euclid \cite{Mellier25}, and the Nancy Grace Roman Space Telescope's High-Latitude Wide-Area Survey (HLWAS) \cite{ROTAC25}, will provide $\gtrsim 10^5$ clusters and their weak lensing signals, with statistical power comparable to that of cosmic shear \cite{Weinberg13, Wu21}.  However, the cosmological constraints from optical clusters have been limited by systematics \cite{To21b, DESY3CL3x2pt, Sunayama24}.  Currently, one key systematic is projection effects---the inclusion of galaxies along the line of sight as cluster members\footnote{
In this paper, we use cluster members to refer to all galaxies identified as members based on photometry.  {\em True} members refer to galaxies physically associated with the main dark matter halo of the host cluster, while {\em projected} members refer to galaxies physically unassociated with this halo.  Observationally, we cannot split true and projected members using 3D cluster radii. Instead, we use two-component fitting to separate the redshift distributions of true and projected members.} \cite{Erickson11, Farahi16, Sunayama20, Wu22, Sunayama23, Cao25}.  Projection effects can lead to biased cluster mass calibration and thus biased cosmological parameters \cite{DESY1CL, Nde26}.

To calibrate cluster projection effects, previous studies have relied on the spectroscopic redshifts of cluster members \cite{Rozo15RM4, Sohn18, Rines18, Capasso19, Myles21projection, Wetzell22, Myles25}.  Such studies require spectra for a representative sample of cluster members.  For the SDSS \redmapper sample of approximately 3000 clusters, 30\% of the members have spectroscopic redshifts from archival data \cite{Rozo16redMaGiC, Myles21projection}. For the DES \redmapper sample, approximately 100 clusters have more than 15 member spectra \cite{Wetzell22}.  The HeCS-omnibus sample includes approximately 200 clusters with a total of 50K member spectra \cite{Sohn20}.  Indeed, it has become increasingly difficult to substantially increase the spectroscopic sample size to keep pace with the growing cluster sample.  Meanwhile, large-area spectroscopic surveys have been taking tens of millions of spectra for the purpose of measuring galaxy clustering; these surveys include BOSS/eBOSS \cite{Dawson13BOSS, Dawson16eBOSS}, DESI \cite{DESI}, Euclid \cite{Mellier25}, 4MOST \cite{4MOST}, and Roman \cite{ROTAC25}.  By construction, these surveys have few spectra in cluster regions because spectrographs cannot take many spectra within a small sky region in a single observation. Nevertheless, they tend to have substantial overlap in area and redshift with cluster samples.

How do we leverage these large spectroscopic surveys to solve the projection effect problem of optical cluster cosmology?  We borrow ideas from the clustering-redshift method (clustering-$z$ hereafter; also known as cross-correlation redshifts, \cite{Newman08, MatthewsNewman10, Schmidt13, Menard13, McQuinnWhite13, Choi16, Morrison17, Johnson17, Davis18, vandenBusch20, Gatti22WZ, Cawthon22, dAssignies25Euclid, dAssignies25DES}), cross-correlating cluster members with a noncluster-targeted spectroscopic sample.  To calibrate the redshift distribution of a photometric sample (often called the {\em unknown} sample), one cross-correlates it with a spectroscopic sample (often called the {\em reference} sample) at a known redshift. A stronger correlation indicates that the two samples occupy similar redshifts.  This cross-correlation, measured in a series of reference redshift bins, allows us to infer the true redshift distribution of the unknown sample.

In our study, the unknown samples are cluster members, while the reference samples are selected from an overlapping spectroscopic survey {\em not} targeting clusters.   We demonstrate our method using the Cardinal simulation, with clusters identified by the \redmapper algorithm \cite{Rykoff14}.  Cluster members are cross-correlated with two distinct spectroscopic samples, luminous red galaxies (LRGs, \citep[][]{Eisenstein01, Zhou23LRG}) and emission line galaxies (ELGs, \citep[][]{Raichoor17, Raichoor23ELG}).  We show that this method can recover the redshift distribution of members and the parameters describing the projected component.    
Although we focus on \redmapper, our method can be used to assess the projection effects of other cluster finders \cite{CAMIRA, WaZP, Grishin23, WenHan24, Yantovski-Barth24}.
Given that our method does not require follow-ups targeting specific cluster members, we envision that it can be used to compare cluster finders applied to the same sky area with spectroscopic coverage.

Since we are interested in the {\em relative} redshift distribution of members with respect to their host clusters, our approach differs from conventional clustering-$z$ implementations.   We bin cluster members based on their host clusters' redshifts instead of individual galaxies' photometric redshifts.

We first put clusters in {\em narrow} redshift bins and count pairs between cluster members and spectroscopic galaxies, in a series of spectroscopic redshift bins.  The resulting pair counts vs.~redshifts are too noisy to infer redshift distributions.  To reduce the noise, we combine clusters in several narrow bins into a {\em wide} redshift bin to stack the pair counts.  With the stacked pair counts, we derive the average redshift distribution of the cluster members.  We then fit a double-Gaussian model to the redshift distribution to derive the projection effect parameters.

This paper is organized as follows. We describe the Cardinal simulation in Sec.~\ref{sec:cardinal} and review the basics of the clustering-$z$ method in Sec.~\ref{sec:basics}.  Section~\ref{sec:implementation} describes our implementation, and Sec.~\ref{sec:results} presents our results.  We compare our work with previous studies and discuss potential systematics in Sec.~\ref{sec:discussion}. Section~\ref{sec:summary} provides the summary and outlook.

\section{The Cardinal simulation}\label{sec:cardinal}

Cardinal \cite{To23Cardinal} is a mock galaxy catalog generated by applying the Addgals algorithm \cite{Wechsler21} to $N$-body simulations.  The Addgals algorithm first uses a high-resolution, small-volume simulation to establish the luminosity-mass and luminosity-density relations of galaxies. It then uses these relations to assign galaxies to low-resolution, large-volume $N$-body simulations with volumes comparable to survey volumes. The algorithm assigns central halo velocities to central galaxies and dark matter particle velocities to satellite galaxies.  The spectroscopic redshift of each galaxy includes these peculiar velocities.

Cardinal is uniquely suitable for our study for two reasons.  First, it has realistic cluster abundances and member populations due to the improved treatment of orphan galaxies.  Second, Cardinal reproduces the galaxy colors of the DES data, enabling us to create realistic spectroscopic LRGs and ELGs. 
We use the light cone designed to mimic the DES data, with a sky coverage of 4,946 deg$^2$.  The underlying $N$-body simulations have a mass resolution of $3.3\times10^{10}~\hiMsun$ for $z<0.315$ and $1.6\times10^{11}~\hiMsun$ for $z>0.315$. They are based on a flat $\Lambda \rm CDM$ cosmology: $\Omega_{\rm m}=0.286$, $\Omega_{\rm b}=0.047$, $\sigma_{8}=0.82$, $n_{\rm s} = 0.96$, and $h=0.7$.  We use these parameter values when converting redshifts to distances.

\subsection{Cardinal's \redmapper clusters}

The \redmapper \cite{Rykoff14} cluster finder is based on the galaxy red sequence---a tight color-magnitude relation for galaxies in clusters.   It finds clusters as overdensities not only on the sky but also in color-magnitude space.  While it does not explicitly use photometric redshifts, the red-sequence method provides photometric redshifts for red galaxies.  The algorithm starts with a modest spectroscopic sample to obtain an initial red-sequence model (the color-magnitude relation as a function of redshift). It then iteratively calibrates the red-sequence model and identifies clusters.   Based on the red-sequence model, \redmapper assigns each red galaxy a photometric redshift $\zred$.   Each red galaxy has the potential to be a central galaxy, and its probability of being a central galaxy is calculated based on its $\zred$, luminosity, and local galaxy density.   
For each central galaxy candidate, \redmapper finds red-sequence galaxies around it, using an iterative 2D aperture, $R_\lambda = (\lambda / 100)^{0.2}~\hiMpc$, and a luminosity threshold $0.2L_*$.
It then assigns each galaxy a membership probability $\pmem$ based on its color, magnitude, and distance to the central galaxy.  The richness $\lambda$ of a cluster is the sum of its members' $\pmem$.   
The cluster redshift $\zlam$ is obtained by jointly fitting all members to the red-sequence model.  For the cluster redshift, $\zcl$, we use the median $\zspec$ of members with $\pmem \geq 0.7$ rather than $\zlam$, assuming that each observed cluster has a few members with available spectroscopic redshifts. For the clustering-$z$ calculation, we use all galaxies provided by the member catalog without setting a $\pmem$ threshold.
We use the volume-limited \redmapper catalog and focus on clusters with $0.3 \le \zcl < 0.75$ and $\lambda \ge 20$.  This selection includes 11,909 clusters and 733,405 members.

\subsection{Cardinal's spectroscopic samples}

We use two distinct spectroscopic galaxy samples to demonstrate our method: LRGs and ELGs. The former tend to be elliptical galaxies, some of which reside in clusters, while the latter tend to be star-forming galaxies and generally do not reside in galaxy clusters.  The sample sizes and binning are summarized in Table~\ref{tab:binning}.

\subsubsection{LRGs}

LRGs are massive elliptical galaxies with negligible star-forming activity and prominent Balmer (4000\AA) breaks.   Their high intrinsic luminosity allows one to construct volume-limited samples across a wide redshift range, and the Balmer break feature leads to accurate photometric redshifts and thus high spectroscopic success rates.   In this work, we select LRGs from the \redmagic sample \cite{Rozo16redMaGiC} from Cardinal.  The \redmagic galaxies are selected using the red-sequence model calibrated by \redmapper; that is, the red-sequence model of \redmapper members is used to estimate the photometric redshifts of noncluster galaxies.  Compared with \redmapper members, \redmagic galaxies are selected to be brighter and to have a smaller scatter about the red sequence (quantified by $\chi^2_{\rm color}$); therefore, they have more accurate photometric redshifts.   Although the \redmagic galaxy sample includes some \redmapper members, most \redmagic galaxies are not \redmapper members.  Therefore, we expect that the cross-correlation between \redmapper members and \redmagic galaxies is dominated by projection effects.

To mimic the spectroscopic LRG sample, we subsample the photometric \redmagic sample to match a comoving density of $3.0 \times 10^{-4}~h^{3}~\mathrm{Mpc}^{-3}$, similar to that of the SDSS LOWZ+CMASS samples \cite{Reid16} and the DESI LRG sample \cite{Zhou23LRG}.  Since \redmagic is based on the red-sequence model of the \redmapper sample, it is limited to $z < 0.75$.

\subsubsection{ELGs}

ELGs are characterized by their [{\sc O ii}] doublet or H$\alpha$ emission lines and tend to be star-forming galaxies. These emission lines enable spectroscopic redshift measurements with relatively short exposure times, especially at high redshifts.  To select ELG targets from an imaging survey, one often uses a luminosity cut in a blue band and a box cut in color-color space optimized for a given redshift range.

We choose ELGs as our reference sample for two reasons. First, we expect that Euclid \cite{Mellier25} and Roman \cite{Wang22} will provide large samples of ELGs.  As Euclid and Roman detect clusters at higher redshifts, cross-correlating clusters and ELGs will become increasingly valuable.  Second, ELGs preferentially reside in filaments and low-density environments rather than in clusters \cite{Gonzalez-Perez20}, making them an extreme case of noncluster-targeted galaxies. Because ELGs have little 3D overlap with clusters, the cross-correlation between ELGs and clusters is dominated by projection effects.

From the Cardinal simulation, we select ELGs following the criteria used to select eBOSS targets from DECaLS \cite{Raichoor17},
\begin{equation}
\begin{aligned}
    21.825 &< g < 22.825, \\
    -0.068~(r-z)+0.457 &< g-r < 0.112(r-z) +0.773, \\
    0.218~(g - r) + 0.571 &< r - z < -0.555~(g - r) + 1.901.
\end{aligned}
\end{equation}
The $g$-band cut selects star-forming galaxies.  This selection is optimized for ELGs at $z = 0.8$, which is higher than the redshifts of our clusters.  This redshift mismatch is not a concern because we are interested in a wide range of spectroscopic sample selections. Compared with LRGs, our ELGs have a lower comoving number density of $5.0\times10^{-5} ~h^{3}{\rm Mpc}^{-3}$ and span a wider redshift range $0\le z < 1$.

\section{Background of the clustering-$z$ method}\label{sec:basics}

We summarize key equations of the clustering-$z$ method following the notation in \cite{Davis18}.  In a clustering-$z$ analysis, one cross-correlates a sample with uncertain redshifts (the unknown sample, indicated by subscripts u) with a sample with precise redshifts (the reference sample, indicated by subscripts r); a strong correlation indicates that the two samples are at similar redshifts.

We use $\phi(z)$ to denote the redshift distribution of galaxies, normalized such that $\int\phi(z)\dd z = 1$.  Let us assume that the redshift distribution of the reference sample is described by a Dirac delta function, $\phir(z) = \delta_{\rm D}(z - \zr)$.  The angular cross-correlation function between the two galaxy samples is given by
\begin{equation} 
    \wur(\theta, \zr) \approx \phiu(\zr) \bu(\zr) \br(\zr) w_\mathrm{mm}(\theta, \zr) ,
\label{eq:wur}
\end{equation}
where $b(z)$ is the linear galaxy bias, and $w_\mathrm{mm}(\theta, \zr)$ is the angular matter correlation function \cite{Simon07}.  We ignore the effects of nonlinear bias, magnification, and redshift-space distortion, which will be discussed in Sec.~\ref{sec:discussion}.

To derive $\phiu(z)$ from $\wur(\theta, \zr)$, we need to model the evolution of other terms.  
We rewrite Eq.~(\ref{eq:wur}) as
\begin{equation}
    \wur(\theta,\zr) \approx \phiu(\zr)  f(\theta, \zr) \wmm(\theta,z=0), 
\end{equation}
where 
$f(\theta, \zr) = \bu(\zr)\br(\zr) G(\theta,\zr)$ captures all the redshift dependence, and $G(\theta,\zr) = \wmm (\theta, \zr)/\wmm(\theta, z=0)$ is the growth factor with possible scale dependence.

Since we are interested in the overall clustering amplitude, we integrate $\wur(\theta,\zr)$ over a weight function $W(\theta) \propto \theta^{-1}$, normalized to unity \cite{Menard13}:
\begin{equation}
\begin{aligned}
    \wurhat(\zr) &\equiv
    \int_{\thetamin}^{\thetamax} \dd \theta W(\theta) \wur(\theta, \zr) \\
    &= \phiu(\zr) \int_{\thetamin}^{\thetamax} \dd\theta W(\theta) f(\theta,\zr) \wmm(\theta,z=0) \\
    &\propto \phiu(\zr) . 
    \label{eq:wurhat_theory}
\end{aligned}
\end{equation}
The last step assumes that the redshift dependence of the integral is weak and can be ignored. In Sec.~\ref{sec:results}, we will show that this assumption leads to unbiased results.

Alternatively, we can apply $W(\theta)$ to pair counts instead of $\wur(\theta,z)$ to reduce noise \cite{Davis18}.  For example, using the Davis-Peebles estimator \citep{DavisPeebles83}, we have
\begin{equation} 
    \wurhat(\zr)  = \frac{\int \dd\theta W(\theta) \Du\Dr(\theta, \zr)}{\int \dd\theta W(\theta) \Du\Rr(\theta, \zr)} - 1, 
\end{equation} 
where $\Du\Dr$ denotes the pair counts of unknown data and reference data, and $\Du\Rr$ denotes the pair counts of unknown data and reference randoms. We omit the factors of total counts in this equation.  With this estimator, the assumption of $\wurhat\propto \phiu(\zr)$ still holds \cite{Davis18}.  Reference~\cite{dAssignies25Euclid} shows that the two weighting schemes give consistent results.  As described below, because our unknown sample is sparse, weighting pair counts improves the signal-to-noise ratio.

\section{Implementation}
\label{sec:implementation}

We now describe our implementation of the clustering-$z$ method for cluster members, which differs from conventional implementations in several ways. We use three types of redshift bins: reference bins, narrow unknown bins, and wide unknown bins.  We first put our unknown galaxies in {\em narrow} bins of $\Delta\zcl=0.0025$ and count the pairs of unknown and reference galaxies.  To improve the signal-to-noise ratio, we combine the pair counts from several adjacent narrow bins to form a {\em wide} bin of $\Delta\zcl=0.05$. Table~\ref{tab:binning} summarizes our binning scheme and the sample sizes in these bins.  Our implementation is detailed as follows:

\begin{table}
\caption{Redshift binning for clustering-$z$ measurements and the corresponding sample sizes.  The sky coverage is 4,946 deg$^2$.}
\label{tab:binning}
\renewcommand{\arraystretch}{1.25}   
\begin{ruledtabular}
\begin{tabular}{l|c|c}
Unknown sample & Clusters & Members  \\
\colrule
Redshift range $\zcl \in [0.3,\,0.75)$ & 12,000 & 730,000 \\
Narrow bin     $\Delta \zcl = 0.0025$  & 7--140 & 270--11,000\\ 
Wide bin       $\Delta \zcl = 0.05$    & 360--2,300 & 17,000--170,000\\
\end{tabular}
\end{ruledtabular}
\smallskip
\begin{ruledtabular}
\begin{tabular}{l|c|c}
Reference samples & LRGs & ELGs \\
\colrule
Redshift range  & $\zr\in[0.15,\,0.75)$ & $\zr\in[0,\,1)$ \\
Total galaxy number & $9.6\times10^5$ &  $1.1\times10^6$ \\
Number density & $3.0\times10^{-4} ~h^{3}{\rm Mpc}^{-3}$ & $5.0\times10^{-5} ~h^{3}{\rm Mpc}^{-3}$\\
Reference bin size & $\Delta\zr=0.005$ & $\Delta\zr=0.005$ \\
Galaxies per bin  & 1,200--16,000 & 430--18,000 \\
\end{tabular}
\end{ruledtabular}
\end{table}

\medskip

{\bf Redshift binning.} 
We choose our bin sizes to be comparable to or smaller than the velocity dispersions of Cardinal clusters, which we estimated to be 750--1500 $\rm km~s^{-1}$ ($\Delta z_{\rm v}$ = 0.0033--0.0086, see Appendix~\ref{app:veldisp}).  We divide our unknown sample (cluster members) into narrow bins of $\Delta\zcl=0.0025$, which is smaller than $\Delta z_{\rm v}$.  On the other hand, we divide the reference sample (either LRGs or ELGs) into bins of $\Delta\zr=0.005$, which is comparable to $\Delta z_{\rm v}$.   

\medskip

{\bf Counting pairs in narrow bins.}
For a given reference bin centered on $\zr$ and a narrow unknown bin, we count u-r pairs as a function of $\theta$, $\Du\Dr(\theta, \zr)$.  We then average over $\theta$ using $W(\theta) \propto \theta^{-1}$,
\begin{equation}
    \Du\Dr(\zr) \propto \sum_{\theta~\rm bins} ~ \Du\Dr(\theta, \zr) W(\theta).
\end{equation}
We perform pair counting using {\tt corrfunc} \cite{Corrfunc} and use 8 $\theta$ bins logarithmically spaced between 0.1 and 10 physical Mpc at each $\zr$ \cite{Davis18}.   Using the same procedure, we compute $\Du\Rr(\zr)$, the angle-averaged pair counts of unknown galaxies and reference randoms. The random catalog is 50 times larger and is generated using the shuffled method \cite{Yang20}.

\medskip

{\bf Combining narrow bins.}
We shift the pair counts $\Du\Dr(\zr)$ and $\Du\Rr(\zr)$ to center on each bin's central $\zcl$. We then combine the pair counts from 20 adjacent narrow bins into a wide bin of $\Delta\zcl=0.05$.  We use $\DuDrhat(z')$ and $\DuRrhat(z')$ to denote the shifted and combined pair counts, where $z'=\zr-\zcl$. (We use the notation $z'$ instead of $\Delta z$ to avoid confusion with bin sizes.)
We then use the Davis-Peebles estimator to calculate the correlation amplitude \citep{DavisPeebles83}
\begin{equation} 
    \wurhat(z') = \frac{\DuDrhat(z')N_{\Rr}}{\DuRrhat(z')N_{\Dr}} - 1, 
\end{equation} 
where $N_{\Dr}$ denotes the number of reference data points, and $N_{\Rr}$ denotes the number of reference randoms. 

\medskip

{\bf Deriving redshift distributions.}
For each $\zcl$, we measure $\wurhat(\zr-\zcl)$ in the range $|\zr-\zcl| \le 0.165$ (67 bins).  We assume $\phiuhat(z) \propto \wurhat(z)$, ignoring the evolution of galaxy biases and the growth factor [Eq.~(\ref{eq:wurhat_theory})].  We normalize the $\wurhat(z)$ to derive the redshift distribution estimator,
\begin{equation}
    \phiuhat(z') = \frac{\wurhat(z')}{\int  \wurhat(z)\dd z~}.
\end{equation}

\medskip

{\bf Calculating jackknife covariance matrices.}
We divide the survey footprint into 100 equal-area patches using {\tt kmeans} \cite{Jarvis04treecorr}; each patch spans approximately 110 Mpc at $z=0.3$.  The jackknife samples are constructed by omitting one patch at a time.  We calculate the $\phiuhat(z')$ for each jackknife sample and calculate the covariance matrices from them.

\medskip

Our end products are the redshift distribution estimators of $\phiuhat(z'=\zr-\zcl)$ in wide bins of $\Delta \zcl = 0.05$, as well as their covariance matrices.
Each $\phiuhat$ is measured with a resolution $\Delta \zr = 0.005$.  We will drop the subscript of $\zr$ from now on.

\begin{figure*}
\centering
\includegraphics[width=2\columnwidth]{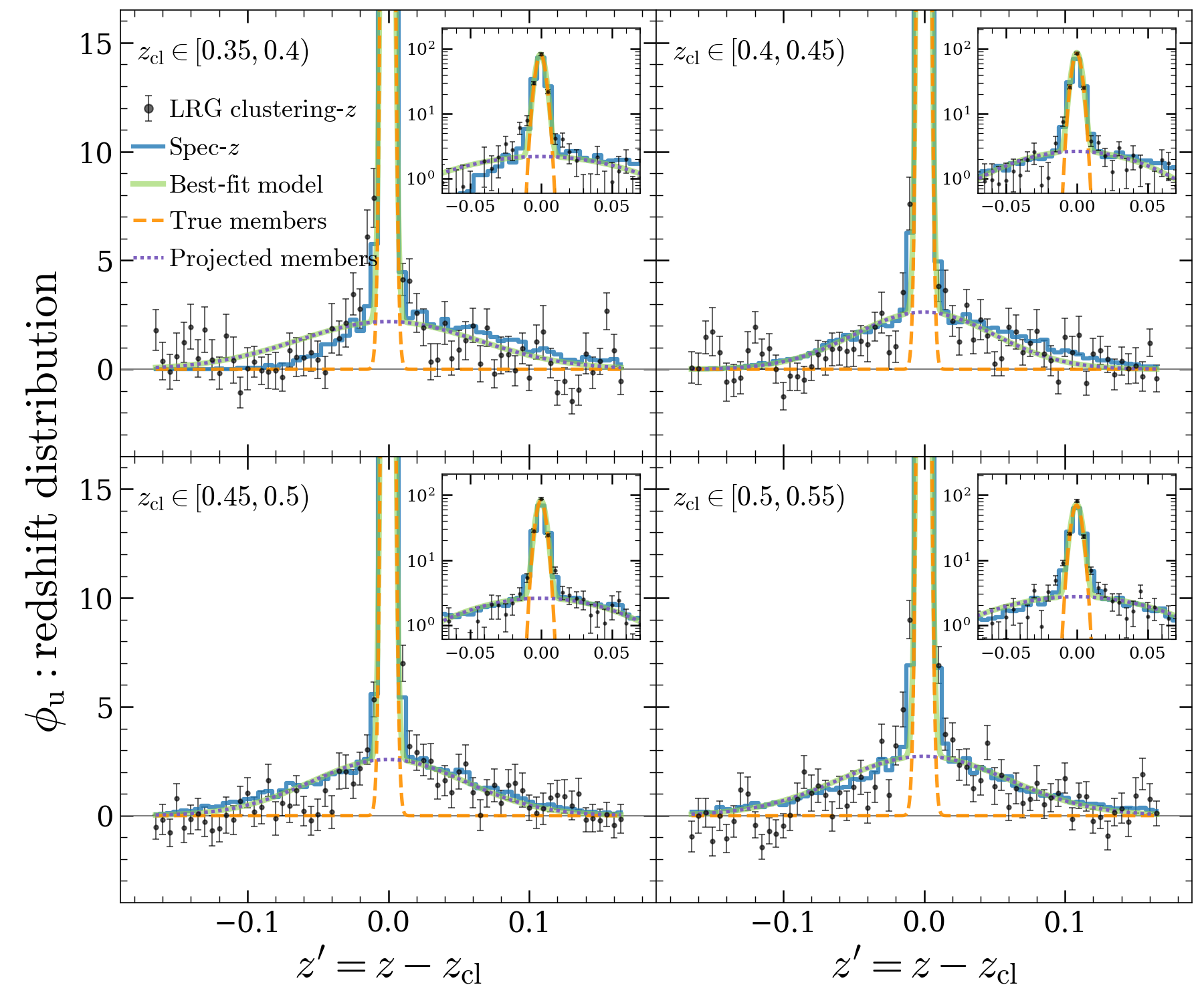}
\caption{Validation of our clustering-$z$ method (cross-correlating cluster members and LRGs) using the Cardinal simulation.  The four panels correspond to four cluster redshift ($z_{\rm cl}$) bins. In each panel, the blue histogram shows the spectroscopic redshift ($z_{\rm spec}$) distribution of cluster members, while the black points with error bars show our clustering-$z$ measurements for the same galaxies.  The agreement shows that the clustering-$z$ method is viable for calibrating the projection effects of cluster members.  In each panel, we fit a double-Gaussian model to $\phiuhat$ to describe true members (orange dashed curves) and projected members (gray dotted curves).  The green curves are the sum of the two Gaussian components. The insets show the central peaks on logarithmic scales.
} 
\label{fig:cardinal_LRG}
\end{figure*}

\begin{figure*}
\centering
\includegraphics[width=2\columnwidth]{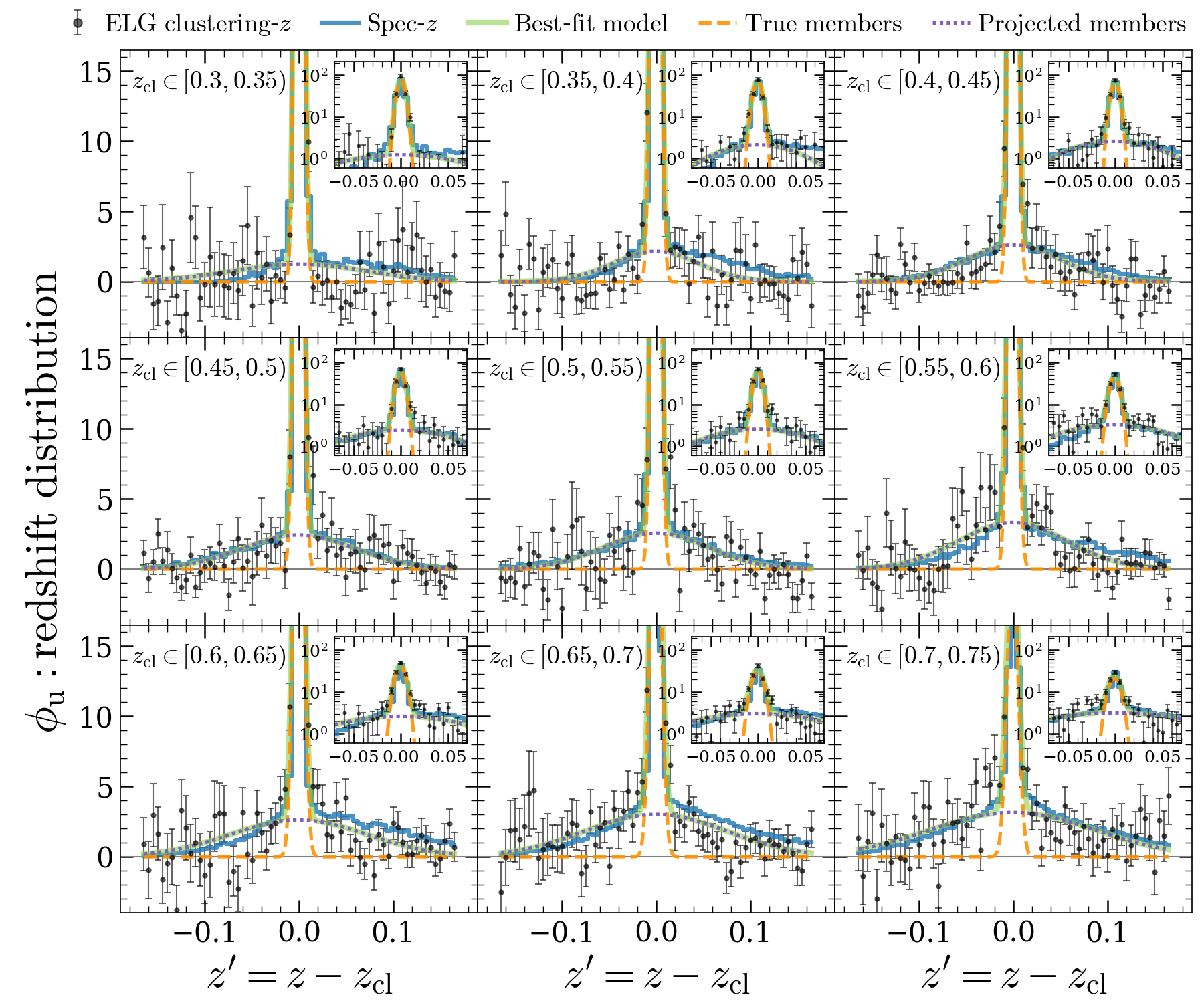}
\caption{Similar to Fig.~\ref{fig:cardinal_LRG}, but using ELGs as the reference sample.  The ELG sample has a wider redshift coverage but a lower number density, resulting in noisier clustering-$z$ measurements.} 
\label{fig:cardinal_ELG}
\end{figure*}

\begin{figure*}
\centering
\includegraphics[width=2\columnwidth]{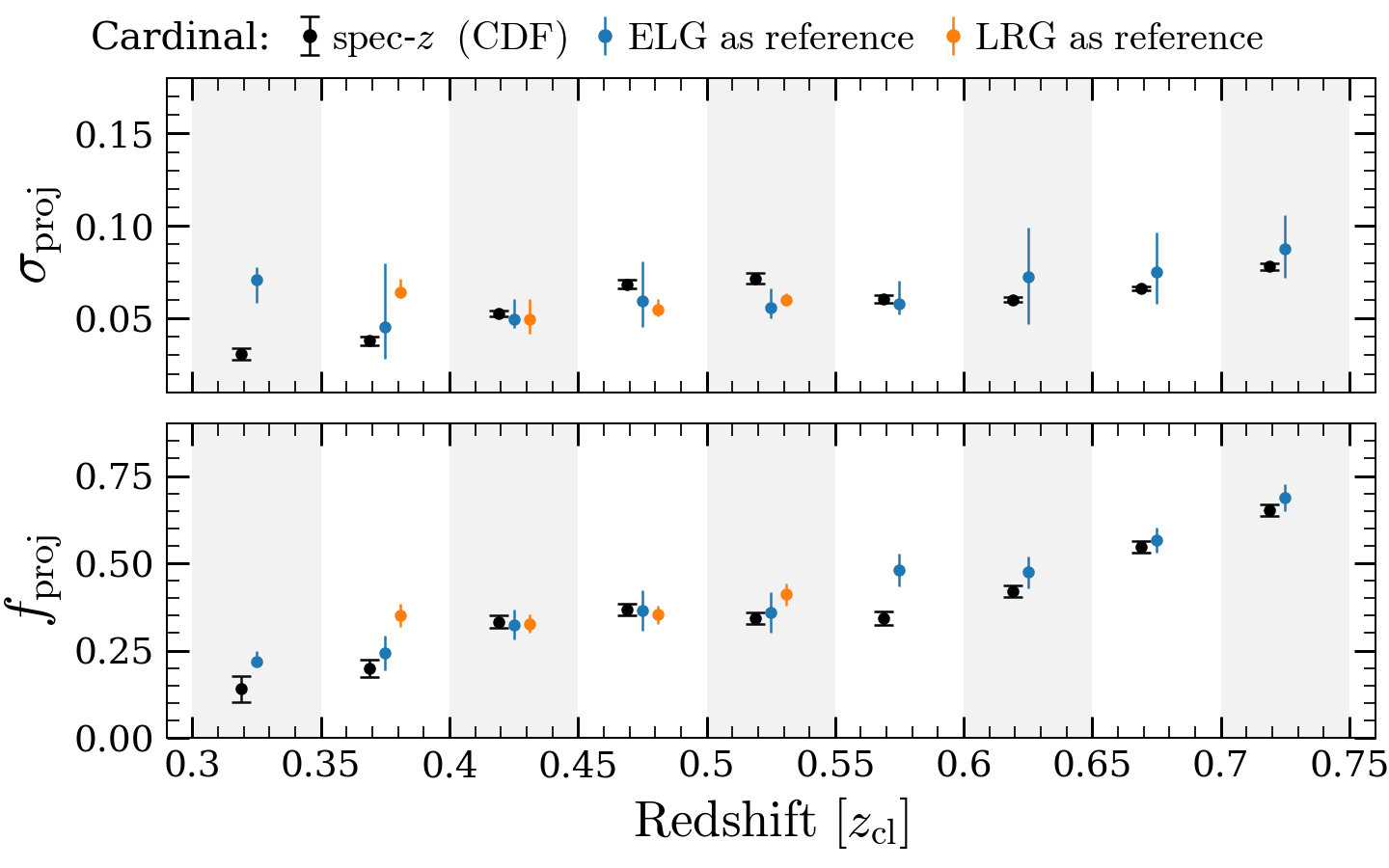}
\caption{Projection effect parameters derived from the clustering-$z$ measurements using LRGs (orange) and ELGs (blue), compared with those derived from spectroscopic redshifts (black).  Both reference samples recover the spectroscopic results, indicating that our method is insensitive to reference sample selection.} 
\label{fig:cardinal_fproj}
\end{figure*}

\section{Results} \label{sec:results}

We now present the redshift distributions of cluster members and the projection effect parameters derived using the clustering-$z$ method. 

\subsection{Redshift distributions}

Figure~\ref{fig:cardinal_LRG} presents the redshift distribution estimator $\phiuhat(z-\zcl)$ for cluster members derived using LRGs as the reference sample (points with error bars).  Each panel corresponds to a wide $\zcl$ bin of $\Delta\zcl=0.05$. The insets zoom in on the central peaks of the distribution.  We can see two components: the central peak corresponds to galaxies physically associated with the clusters, while the broad wings correspond to projected members.  In the wings, we see some negative values due to the low number of galaxy pairs.

The blue histograms show the spectroscopic redshift $\zspec$ distribution of the same cluster members.   We can see that the clustering-$z$ method recovers the $\zspec$ distribution.  At the lowest redshift bin, the $\zspec$ distribution is asymmetric around $\zcl$ due to increased redshift uncertainties associated with the filter transition at $z\approx0.35$ (the 4000\AA\xspace break passing through filter gaps \cite{Rykoff14}).  Because Cardinal assigns DES galaxies' magnitudes to simulated galaxies, its photometry is affected by the filter transition.

Figure~\ref{fig:cardinal_ELG} presents the analogous calculation using ELGs as the reference sample.  The clustering-$z$ method again recovers the spectroscopic redshift distribution.  The ELG sample has a wider redshift coverage than LRGs but a lower number density, resulting in noisier $\phiuhat$ measurements.  Comparing the LRG and ELG results, we can see that our method is insensitive to sample selection.

\subsection{Quantifying projection effects}

We model the redshift distribution of cluster members using a double-Gaussian function \cite{Myles21projection}
\begin{equation}
    \phiu^{\rm model}(z') =
    (1-\fproj)\,\mathcal{N}(z' \mid \mu, \sigmacl^2)
    + \fproj\,\mathcal{N}(z' \mid \mu, \sigmaproj^2),
\label{eq:double_gaussian}
\end{equation} 
where $z' = z - \zcl$, and $\mu$ is the offset of the peak from zero.  The first Gaussian distribution describes true members, while the second describes projected members.  The $\fproj$ parameter describes the fraction of projected members, and $\sigmaproj$ describes their redshift extent.  This $\fproj$ is derived from a statistical fit and may differ from the fraction of galaxies outside the halo's virial radius \cite{Cao25}.

We fit this model to our measured $\phiuhat$ by minimizing the $\chi^2$, defined with the jackknife covariance matrix $\mathsf{\Sigma}$, 
\begin{equation}
    \chi^2 = (\phiuhat - \phiu^{\rm model})^{\rm T} {\mathsf{\Sigma}}^{-1}(\phiuhat - \phiu^{\rm model}).
\end{equation}
We adopt uniform priors for the model parameters
$\fproj \sim \mathcal{U}(0.01, 0.8)$,
$\mu \sim \mathcal{U}(-0.01, 0.01)$,
$\sigmacl \sim \mathcal{U}(0.001, 0.01)$, and
$\sigmaproj \sim \mathcal{U}(0.01, 0.1)$.
We run Markov chain Monte Carlo using the ensemble sampler \texttt{emcee} \citep{emcee} with 200 walkers for 200,000 to 300,000 steps to obtain the posterior distributions of the parameters. In each panel of Fig.~\ref{fig:cardinal_LRG} and Fig.~\ref{fig:cardinal_ELG}, we show the best-fit model and its two components.

Figure~\ref{fig:cardinal_fproj} presents $\sigmaproj$ and $\fproj$ as functions of redshift. The orange and blue points correspond to the parameters derived from LRGs and ELGs, and the error bars show the 68\% intervals of the posterior distribution.  The black points correspond to fitting the cumulative distribution function of $\zspec$ and are considered the true values.  We do not show $\mu$ and $\sigmacl$; $\mu$ is consistent with zero ($|\mu|< 10^{-3}$) and $\sigmacl$ is affected by the bin sizes.  We have checked that the parameters describing the projected component are insensitive to bin sizes.

As can be seen, the projection effect parameters derived from both clustering-$z$ measurements are consistent with those derived from spectroscopic redshifts.  The consistency between LRG and ELG results indicates that our method is insensitive to the spectroscopic sample selection.  The LRG sample has a limited redshift range because it relies on \redmapper's red-sequence calibration.  When available, the LRG results have smaller error bars due to their higher number densities.

The $\fproj$ parameter increases from 0.14 ($\zcl=0.3$) to 0.65 ($\zcl=0.75$), confirming that the projection fraction increases with redshift \cite{Costanzi19projection, Grandis25}.  Our $\fproj$ at $\zcl=0.3$ is lower than the values for SDSS \redmapper derived from $\zspec$:
$\fproj=0.265$ at $z\approx0.1$ \cite{Myles21projection} and 
$\fproj=0.340$ at $z\approx0.2$ \cite{Myles25}.

On the other hand, $\sigmaproj$ shows a gradual, nonmonotonic increase from 0.03 ($\zcl=0.3$) to 0.08 ($\zcl=0.75$), corresponding to 78 $\hiMpc$ to 160 $\hiMpc$ assuming Cardinal's cosmology.  The values at higher redshift are higher than the SDSS results from \cite{Myles21projection}, $0.032$ (corresponding to 8689 ${\rm km~s^{-1}}$ and $90~\hiMpc$).

As shown in Fig.~\ref{fig:cardinal_LRG} and Fig.~\ref{fig:cardinal_ELG}, for low-redshift clusters, the member redshift distribution is asymmetric.  As a result, a double-Gaussian model does not always provide a good fit.  In Fig.~\ref{fig:cardinal_fproj}, small discrepancies between $\zspec$ results and clustering-$z$ results are partly driven by this asymmetry.  Our fiducial calculation uses all \redmapper members without a $\pmem$ threshold.  We have tested that using different $\pmem$ thresholds gives similar results.

\section{Discussion}\label{sec:discussion}

In this section, we first compare our projection effect results with those in the literature and discuss potential systematics in our clustering-$z$ approach.

\subsection{Comparison with previous studies on projection effects}

\citet{Sunayama23} uses the cluster-galaxy cross-correlation to constrain projection effects, focusing on the SDSS \redmapper (red-sequence Matched-filter Probabilistic Percolation) clusters and the LOWZ spectroscopic galaxy sample. They calculate the projected correlation function $w_{\rm p,cg}(r_{\rm p})$ using different line-of-sight integration lengths $\pi_{\rm max}$.  They find that the ratio $w_{\rm p,cg}/w_{\rm p,gg}$ increases with $\pi_{\rm max}$ until $\pi_{\rm max} \approx 150 ~ \hiMpc$, which provides an upper limit on the distance of projected members.  Our results are consistent with their findings.

\citet{Lee25} generate mock cluster catalogs using different projection effect models and compare them with several properties of SDSS \redmapper clusters at $z=0.1$: cluster counts, lensing, spectroscopic redshift distribution of members, and richness remeasured on a redshift grid. They find that these observables agree with a Gaussian projection model with $\Delta\chi = 110 ~\hiMpc$.  This projection model is consistent with our results and with \cite{Myles21projection}.

\citet{Myles25} use SDSS \redmapper clusters at $z \lesssim 0.22$ and the spectroscopic redshifts of their members to quantify projection effects.  The authors show that the redshift distribution of projected members tends to be asymmetric, with more projected members behind the clusters than in front of them. The asymmetry is more significant for faint galaxies ($L < 0.5 L_*$).  They reason that galaxies behind clusters tend to be fainter and have larger redshift uncertainties.  However, in Cardinal, the asymmetry is strongest for $0.35 \le z < 0.4$ and weaker at higher redshifts.  Therefore, in our analysis, the asymmetry is most likely driven by a filter transition: the 4000\AA\xspace break in galaxy spectra shifts from the $g$ filter to the $r$ filter at $z\approx0.35$. The filter transition is known to lead to non-Gaussian scatter in photometric redshifts \cite{Rykoff14}.

\subsection{Systematic uncertainties in the clustering-$z$ method}

In this section, we discuss systematic uncertainties that can affect clustering-$z$ measurements, in the context of galaxy clusters. These uncertainties are negligible in our study but warrant further investigation.  

\medskip

{\bf Galaxy bias evolution.}
The accuracy of the clustering-$z$ method relies on knowledge of $\bu(z)$, i.e., the bias evolution of the unknown sample [Eq.~(\ref{eq:wur})]. [The evolution of the reference sample $\br(z)$ is always negligible because we use fine reference bins.]  Previous studies have developed different strategies for calibrating $\bu(z)$, including iteratively solving for the correlation function parameters and redshift distribution \cite{Newman08, MatthewsNewman10}, fitting parametrized models \cite{Davis18, dAssignies25DES}, and measuring $w_{\rm uu}$ in narrow photometric redshift bins \cite{Cawthon22, dAssignies25Euclid}.  In this work, we ignore the evolution of $\bu(z)$ and obtain unbiased results.  Each of our unknown samples spans a redshift range $\Delta z \lesssim 0.2$ (see Fig.~\ref{fig:cardinal_fproj}), and we expect the bias evolution to be negligible in such a range.

\medskip

{\bf Nonlinear galaxy bias.}
Clustering-$z$ studies routinely use small-scale clustering to reduce noise and avoid overlap with 3$\times$2pt analyses. At small scales, galaxy bias is nonlinear, but the linear bias assumption of Eq.~(\ref{eq:wur}) has been shown to be valid \cite{Schmidt13}.  In this work, we assume that any scale dependence of the correlation functions is averaged out by the angular integration in Eq.~(\ref{eq:wurhat_theory}).  The agreement between the clustering-$z$ results and the spectroscopic redshift distribution indicates that the nonlinear galaxy bias does not significantly impact our results.  Recently, Ref.~\cite{dAssignies25Euclid} advocates excluding 1-halo scales to avoid nonlinear effects.  We will investigate the impact of nonlinear galaxy bias in the future.

\medskip

{\bf Magnification.} 
Gravitational lensing magnification leads to correlations between galaxies that are well separated in redshift, contributing to systematic biases of $\wur$.  For example, reference galaxies at $z=0.6$ can be magnified by unknown galaxies at $z=0.3$, leading to a nonzero correlation between them.  The impact of magnification can be modeled analytically or through simulations \cite{Gatti22WZ, Cawthon22, dAssignies25Euclid}.   Our study focuses on cluster regions, which tend to have high magnification.  However, since our unknown and reference samples are relatively close in redshift, we expect magnification to have a negligible impact on our results.  Also, our unknown samples are in fine redshift bins, which reduces the impact of magnification \cite{dAssignies25Euclid}. 

\medskip

{\bf Redshift-space distortions.}
Redshift-space distortions arise because spectroscopic redshifts include galaxies' line-of-sight peculiar velocities.  In the context of clustering-$z$ measurements, a spectroscopic redshift bin includes galaxies from neighboring cosmological redshift bins.  The Cardinal simulation includes this effect, but our formalism [Eq.~(\ref{eq:wur})] does not account for it. Reference~\cite{dAssignies25Euclid} uses simulations to show that peculiar velocities of the reference sample do not significantly affect the clustering-$z$ results.  Although cluster members have larger peculiar velocities than field galaxies do, in our study, they serve as unknown samples and do not impact the spectroscopic binning.

\medskip

{\bf Sample sizes.}
The sample size determines the noise level of clustering-$z$ measurements. Our unknown sample size is smaller than those of typical photometric galaxy samples; for example, Ref.~\cite{Cawthon22} includes 60,000--300,000 galaxies per unknown bin, while our calculation includes 270--11,000 members in each narrow bin of $\Delta\zcl=0.0025$, and 17,000--170,000 members in each wide bin of $\Delta\zcl=0.05$ (360--2,300 clusters per wide bin); see Table~\ref{tab:binning}.  Despite our smaller unknown sample size, we still recover the spectroscopic redshift distribution.  On the other hand, our reference sample size is comparable to that of currently available samples.  We use approximately 1,200--16,000 LRGs and 430--18,000 ELGs per reference bin of $\Delta\zr = 0.005$.  We find that using 50\% and even 30\% of this ELG sample yields consistent, albeit noisier, results.

\section{Summary and outlook}\label{sec:summary}

We have developed a clustering-$z$ approach to calibrate the projection effects of optically selected clusters, using noncluster-targeted spectroscopic surveys.  Instead of directly using the available spectroscopic redshifts of cluster members, we cross-correlate all members with a sparse sample of spectroscopic galaxies selected for galaxy clustering studies, most of which are {\em not} cluster members.

We have validated our method using the \redmapper clusters from the Cardinal simulation.  We use LRGs and ELGs as reference spectroscopic samples; the former have some overlap with cluster members, while the latter have negligible overlap.  
Because clusters have a low number density, we use two levels of cluster redshift binning: (1) narrow bins for calculating the pair counts between members and spectroscopic galaxies, and (2) wide bins for combining clusters at different redshifts.  Using this approach, we have inferred the redshift distribution of cluster members and fitted a double-Gaussian model to it.  We have shown that using either LRGs or ELGs recovers the spectroscopic redshift distribution of \redmapper members and the correct projection effect parameters.

We plan to apply this approach to photometric surveys that have substantial overlap with BOSS and DESI, such as the DECADE survey \cite{Anbajagane25DECADE}.  This approach will also be applicable to several upcoming datasets.  The Euclid mission overlaps with DESI, 4MOST, and BOSS and has its own spectroscopic observations of ELGs via NISP slitless spectroscopy for $z>0.9$ \cite{Mellier25}.  The LSST data will overlap with DESI over at least 4,000 square degrees \cite{LSSTSRD18}.  The HLWAS from Roman will obtain spectra for approximately 2,000 square degrees \cite{Wang22}.

Although we use \redmapper as a case study, we plan to apply our clustering-$z$ approach to clusters identified by other algorithms \cite{CAMIRA, WaZP, Grishin23, WenHan24, Yantovski-Barth24}.  Our approach provides a common framework for assessing projection effects across different cluster-finding methods, independent of the available spectroscopic follow-ups.

\begin{acknowledgments}

We thank the anonymous referee for helpful suggestions.  This study was supported by DOE award DE-SC0010129 and NSF award AST-2509910.  

Computations for this study were performed on the M3 high-performance computing cluster provided by the Office of Information Technology (OIT) and the O'Donnell Data Science and Research Computing Institute of Southern Methodist University. 

\end{acknowledgments}

\section*{Data Availability}

The datasets used in this work are available upon request.

\appendix

\section{Velocity dispersions of \redmapper clusters in Cardinal}
\label{app:veldisp}

\begin{figure}
\centering
\includegraphics[width=\columnwidth]{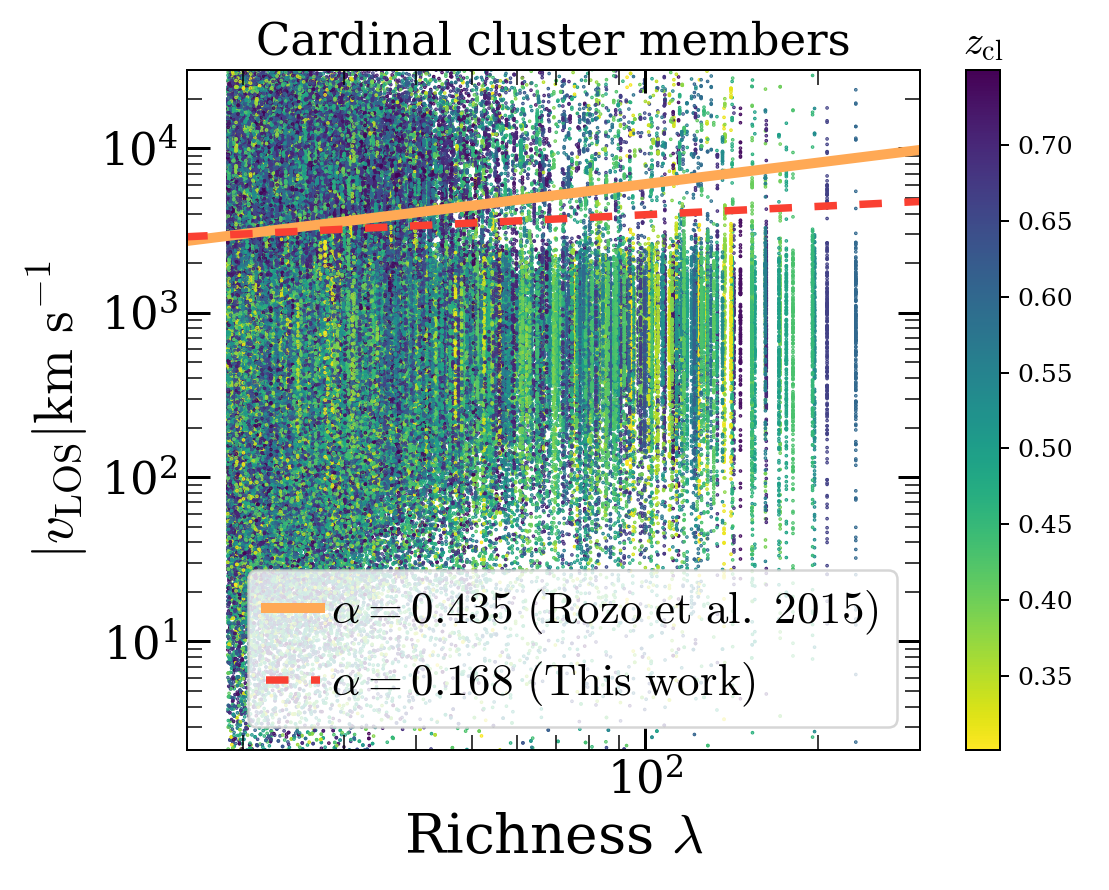}
\caption{Individual \redmapper members' line-of-sight velocities vs.~the richness of their host clusters. We use a maximum-likelihood approach to estimate velocity dispersions vs.~richness.  
The orange line corresponds to the boundary we use to separate true and projected members and is taken from \cite{Rozo15RM4}, while the red dashed line indicates the shallower slope resulting from our fit.
} 
\label{fig:veldisp}
\end{figure}

In this appendix, we estimate the velocity dispersions of \redmapper clusters, which inform our choices of redshift bin sizes.  We need bin sizes comparable to or smaller than the velocity dispersions, yet wide enough to include a sufficient number of pairs.  The usual definition of the velocity dispersion of halos---based on the 3D positions and velocities of galaxies within a halo radius---is not suitable for our purpose, because our true vs.~projected members are separated by two-component fitting using Eq.~(\ref{eq:double_gaussian}) rather than by cuts in 3D distances. Therefore, we adopt a maximum-likelihood approach presented in \cite{Rozo15RM4} to mimic the observational procedure of estimating velocity dispersion.

This maximum-likelihood approach is based on the line-of-sight velocities of all members with respect to their host cluster,
\begin{equation}
    v_{\rm LOS} = c\frac{z_{\rm mem}-\zcl}{1+\zcl}.
\end{equation}
Figure~\ref{fig:veldisp} shows $v_{\rm LOS}$ vs.~$\lambda$ of the cluster members used in this study, with $0.3 \le \zcl < 0.75$.  We apply a velocity cut to exclude projected members with $|v_{\rm LOS}| > (3000 \rm ~km ~s^{-1} )(\lambda/20)^{\alpha}$ with $\alpha=0.435$ (best-fit value from \cite{Rozo15RM4}), shown as the orange line.

We assume that velocity dispersions scale with richness and redshift via
\begin{equation}
    \sigmav(\lambda, \zcl) = \sigma_{\rm p}\bigg(\frac{1+\zcl}{1+z_{\rm p}}\bigg)^{\beta}\bigg(\frac{\lambda}{\lambda_{\rm p}}\bigg)^{\alpha}. 
\end{equation}
The pivot redshift and richness are based on the median values of our sample, $z_{\rm p}=0.524$ and $\lambda_{\rm p}=28.424$.

The likelihood function for individual members' line-of-sight velocities is modeled as
\begin{equation}
    \mathcal{L}=\prod_i \left[p~\mathcal{N}\left(v_{{\rm LOS},i} \mid 0, \sigmav^2(\lambda, \zcl)\right)+\frac{1-p}{2v_{\rm max}}\right],
\end{equation}
where $i$ runs through all cluster members of all clusters, and we use the $\lambda$ and $\zcl$ of each member's host cluster. The first term in the brackets corresponds to physically associated members, and the second term corresponds to projected members.  To compare with \cite{Rozo15RM4}, we rescale the DES-based richness in Cardinal to the SDSS richness using  $\lambda_{\rm SDSS} = 0.92~\lambda_{{\rm DES{\text-}Y3}} + 0.45$ \cite{Wetzell22}.  We choose $v_{\rm max}=6944~{\rm km~s}^{-1}$, which is the maximum velocity below the velocity cut (orange line in Fig.~\ref{fig:veldisp}).

Our best-fit parameters are 
\begin{equation}
    \begin{aligned}
        \sigma_{\rm p} &=957.153 \pm 14.5 ~\rm km~s^{-1}, \\
        \alpha&= 0.168\pm 0.0277, \\
        \beta&= 1.199 \pm 0.0937,\\
        p&= 0.973 \pm 0.00707.
    \end{aligned}
\end{equation}
We convert $\sigmav$ to the redshift difference using $\Delta z_{\rm v} =   (\sigmav / c)(1+ \zcl)$.  For the full cluster sample, $\Delta z_{\rm v}$ ranges from 0.0033 ($747.6~\rm km~s^{-1}$) to 0.0086 ($1505.4~\rm km~s^{-1}$).

Reference~\cite{Rozo15RM4} finds $\alpha = 0.435 \pm 0.020$, $\beta = 0.54 \pm 0.19$, and $p = 0.9163 \pm 0.0042$ for the SDSS \redmapper; our scaling with richness ($\alpha$) is weaker, and our scaling with redshift ($\beta$) is stronger.  This difference is likely because we include a wider redshift range and fainter galaxies than those in \cite{Rozo15RM4}.  On the other hand, Ref.~\cite{Wetzell22} calculates the velocity dispersions of individual DES \redmapper clusters with more than 15 members each, using the biweight and the gapper methods.  They find a population of velocity outliers corresponding to projected members.  They also find lower velocity dispersions than those in \cite{Rozo15RM4}, indicating that the maximum-likelihood approach may overestimate the velocity dispersions.  Since we use these velocity dispersion estimates only to inform the redshift bin choices, any possible overestimation does not affect our results.

\bibliographystyle{apsrev4-2-author-truncate-noarxiv}
\bibliography{master_refs}

@ARTICLE{Yang20,
       author = {{Yang}, Lei and {Jing}, Yi-Peng and {Li}, Zhi-Gang and {Yang}, Xiao-Hu},
        title = "{Toward accurate measurement of property-dependent galaxy clustering: I. Comparison of the V$_{max}$ method and the ``shuffled'' method}",
      journal = {Research in Astronomy and Astrophysics},
         year = 2020,
        month = apr,
       volume = {20},
       number = {4},
          eid = {054},
        pages = {054},
          doi = {10.1088/1674-4527/20/4/54},
archivePrefix = {arXiv},
       eprint = {1912.05976},
 primaryClass = {astro-ph.CO},
       adsurl = {https://ui.adsabs.harvard.edu/abs/2020RAA....20...54Y}
}

@ARTICLE{Maturi25KiDS,
       author = {{Maturi}, M. and {Radovich}, M. and {Moscardini}, L. and {Lesci}, G.~F. and {Castignani}, G. and {Marulli}, F. and {Puddu}, E.~A. and {Romanello}, M. and {Sereno}, M. and {Giocoli}, C. and {Ingoglia}, L. and {Bardelli}, S. and {Giblin}, B. and {Hildebrandt}, H. and {Joudaki}, S.},
        title = "{AMICO galaxy clusters in KiDS-1000: Cosmological sample}",
      journal = {\aap},
         year = 2025,
        month = sep,
       volume = {701},
          eid = {A201},
        pages = {A201},
          doi = {10.1051/0004-6361/202554340},
archivePrefix = {arXiv},
       eprint = {2507.14338},
 primaryClass = {astro-ph.CO},
       adsurl = {https://ui.adsabs.harvard.edu/abs/2025A&A...701A.201M}
}

@ARTICLE{Dawson13BOSS,
       author = {{Dawson}, Kyle S. and {Schlegel}, David J. and {Ahn}, Christopher P. and {Anderson}, Scott F. and {Aubourg}, {\'E}ric and {Bailey}, Stephen and {Barkhouser}, Robert H. and {Bautista}, Julian E. and {Beifiori}, Alessandra and {Berlind}, Andreas A. and {Bhardwaj}, Vaishali and {Bizyaev}, Dmitry and {Blake}, Cullen H. and {Blanton}, Michael R. and {Blomqvist}, Michael and others},
        title = "{The Baryon Oscillation Spectroscopic Survey of SDSS-III}",
      journal = {\aj},
         year = 2013,
        month = jan,
       volume = {145},
       number = {1},
          eid = {10},
        pages = {10},
          doi = {10.1088/0004-6256/145/1/10},
archivePrefix = {arXiv},
       eprint = {1208.0022},
 primaryClass = {astro-ph.CO},
       adsurl = {https://ui.adsabs.harvard.edu/abs/2013AJ....145...10D}
}

@ARTICLE{Dawson16eBOSS,
       author = {{Dawson}, Kyle S. and {Kneib}, Jean-Paul and {Percival}, Will J. and {Alam}, Shadab and {Albareti}, Franco D. and {Anderson}, Scott F. and {Armengaud}, Eric and {Aubourg}, {\'E}ric and others},
        title = "{The SDSS-IV Extended Baryon Oscillation Spectroscopic Survey: Overview and Early Data}",
      journal = {\aj},
         year = 2016,
        month = feb,
       volume = {151},
       number = {2},
          eid = {44},
        pages = {44},
          doi = {10.3847/0004-6256/151/2/44},
archivePrefix = {arXiv},
       eprint = {1508.04473},
 primaryClass = {astro-ph.CO},
       adsurl = {https://ui.adsabs.harvard.edu/abs/2016AJ....151...44D}
}

@ARTICLE{dAssignies25DES,
       author = {{d'Assignies}, W. and {Bernstein}, G.~M. and {Yin}, B. and {Giannini}, G. and {Alarcon}, A. and {Manera}, M. and {To}, C. and {Yamamoto}, M. and {Weaverdyck}, N. and {Cawthon}, R. and {Gatti}, M. and {Amon}, A. and {Anbajagane}, D. and {Avila}, S. and {Becker}, M.~R. and {Bechtol}, K. and {Chang}, C. and {Crocce}, M. and {De Vicente}, J. and {Dodelson}, S. and {Fang}, J. and {Fert{\'e}}, A. and {Gruen}, D. and {Legnani}, E. and {Porredon}, A. and {Prat}, J. and {Rodriguez-Monroy}, M. and {S{\'a}nchez}, C. and {Schutt}, T. and {Sevilla-Noarbe}, I. and {Sanchez Cid}, D. and {Troxel}, M.~A. and {Abbott}, T.~M.~C. and {Andrade-Oliveira}, F. and {Aguena}, M. and {Alves}, O. and {Bacon}, D. and {Blazek}, J. and {Bocquet}, S. and {Brooks}, D. and {Camilleri}, R. and {Carnero Rosell}, A. and {Carrasco Kind}, M. and {Carretero}, J. and {Castander}, F.~J. and {da Costa}, L.~N. and {da Silva Pereira}, M.~E. and {Davis}, T.~M. and {Desai}, S. and {Doel}, P. and {Doux}, C. and {Drlica-Wagner}, A. and {Eifler}, T. and {Elvin-Poole}, J. and {Everett}, S. and {Flaugher}, B. and {Fosalba}, P. and {Frieman}, J. and {Garc{\'\i}a-Bellido}, J. and {Gaztanaga}, E. and {Giles}, P. and {Gutierrez}, G. and {Hinton}, S.~R. and {Hollowood}, D.~L. and {Honscheid}, K. and {Huterer}, D. and {Jain}, B. and {James}, D.~J. and {Kuehn}, K. and {Lahav}, O. and {Lee}, S. and {Marshall}, J.~L. and {Mena-Fern{\'a}ndez}, J. and {Menanteau}, F. and {Miquel}, R. and {Muir}, J. and {Myles}, J. and {Ogando}, R.~L.~C. and {Palmese}, A. and {Paterno}, M. and {Petravick}, P. and {Plazas Malag{\'o}n}, A.~A. and {Raveri}, M. and {Roodman}, A. and {Samuroff}, S. and {Sanchez}, E. and {Sheldon}, E. and {Shin}, T. and {Smith}, M. and {Suchyta}, E. and {Swanson}, M.~E.~C. and {Tarle}, G. and {Thomas}, D. and {Vikram}, V. and {Walker}, A.~R. and {DES Collaboration}},
        title = "{Dark Energy Survey Year 6 results: Clustering redshifts and importance sampling of self-organized-maps n(z) realizations for 3{\texttimes}2pt samples}",
      journal = {\prd},
         year = 2026,
        month = jun,
       volume = {113},
       number = {12},
          eid = {123549},
        pages = {123549},
          doi = {10.1103/6lfl-h6dg},
archivePrefix = {arXiv},
       eprint = {2510.23565},
 primaryClass = {astro-ph.CO},
       adsurl = {https://ui.adsabs.harvard.edu/abs/2026PhRvD.113l3549D}
}

@ARTICLE{dAssignies25Euclid,
       author = {{d'Assignies}, W. and {Manera}, M. and {Padilla}, C. and {Ilbert}, O. and {Hildebrandt}, H. and {Reynolds}, L. and {Chaves-Montero}, J. and {Wright}, A.~H. and {Tallada-Cresp{\'\i}}, P. and {Eriksen}, M. and {Carretero}, J. and {Roster}, W. and {Kang}, Y. and {Naidoo}, K. and {Miquel}, R. and {Altieri}, B. and {Amara}, A. and {Andreon}, S. and {Auricchio}, N. and {Baccigalupi}, C. and {Bagot}, D. and {Baldi}, M. and {Balestra}, A. and {Bardelli}, S. and {Battaglia}, P. and {Biviano}, A. and {Branchini}, E. and {Brescia}, M. and {Camera}, S. and {Capobianco}, V. and {Carbone}, C. and {Cardone}, V.~F. and {Casas}, S. and {Castander}, F.~J. and {Castellano}, M. and {Castignani}, G. and {Cavuoti}, S. and {Chambers}, K.~C. and {Cimatti}, A. and {Colodro-Conde}, C. and {Congedo}, G. and {Conselice}, C.~J. and {Conversi}, L. and {Copin}, Y. and {Courbin}, F. and {Courtois}, H.~M. and {Crocce}, M. and {Da Silva}, A. and {Degaudenzi}, H. and {de la Torre}, S. and {De Lucia}, G. and {Douspis}, M. and {Dupac}, X. and {Ealet}, A. and {Escoffier}, S. and {Farina}, M. and {Faustini}, F. and {Ferriol}, S. and {Finelli}, F. and {Fosalba}, P. and {Fotopoulou}, S. and {Frailis}, M. and {Franceschi}, E. and {Fumana}, M. and {Galeotta}, S. and {George}, K. and {Gillis}, B. and {Giocoli}, C. and {G{\'o}mez-Alvarez}, P. and {Gracia-Carpio}, J. and {Grazian}, A. and {Grupp}, F. and {Holmes}, W. and {Hook}, I.~M. and {Hornstrup}, A. and {Jahnke}, K. and {Jhabvala}, M. and {Joachimi}, B. and {Keih{\"a}nen}, E. and {Kermiche}, S. and {Kiessling}, A. and {Kubik}, B. and {K{\"u}mmel}, M. and {Kunz}, M. and {Kurki-Suonio}, H. and {Lahav}, O. and {Le Brun}, A.~M.~C. and {Ligori}, S. and {Lilje}, P.~B. and {Lindholm}, V. and {Lloro}, I. and {Mainetti}, G. and {Maino}, D. and {Maiorano}, E. and {Mansutti}, O. and {Marcin}, S. and {Marggraf}, O. and {Markovic}, K. and {Martinelli}, M. and {Martinet}, N. and {Marulli}, F. and {Massey}, R. and {Masters}, D.~C. and {Medinaceli}, E. and {Mei}, S. and {Melchior}, M. and {Mellier}, Y. and {Meneghetti}, M. and {Merlin}, E. and {Meylan}, G. and {Mora}, A. and {Moresco}, M. and {Moscardini}, L. and {Neissner}, C. and {Niemi}, S.-M. and {Paltani}, S. and {Pasian}, F. and {Pedersen}, K. and {Pettorino}, V. and {Pires}, S. and {Polenta}, G. and {Poncet}, M. and {Popa}, L.~A. and {Pozzetti}, L. and {Raison}, F. and {Rebolo}, R. and {Renzi}, A. and {Rhodes}, J. and {Riccio}, G. and {Romelli}, E. and {Roncarelli}, M. and {Rossetti}, E. and {Saglia}, R. and {Sakr}, Z. and {Sapone}, D. and {Sartoris}, B. and {Schewtschenko}, J.~A. and {Schneider}, P. and {Schrabback}, T. and {Secroun}, A. and {Sefusatti}, E. and {Seidel}, G. and {Seiffert}, M. and {Serrano}, S. and {Simon}, P. and {Sirignano}, C. and {Sirri}, G. and {Spurio Mancini}, A. and {Stanco}, L. and {Steinwagner}, J. and {Tavagnacco}, D. and {Taylor}, A.~N. and {Teplitz}, H.~I. and {Tereno}, I. and {Tessore}, N. and {Toft}, S. and {Toledo-Moreo}, R. and {Torradeflot}, F. and {Tsyganov}, A. and {Tutusaus}, I. and {Valenziano}, L. and {Valiviita}, J. and {Vassallo}, T. and {Verdoes Kleijn}, G. and {Wang}, Y. and {Weller}, J. and {Zamorani}, G. and {Zucca}, E. and {Bolzonella}, M. and {Burigana}, C. and {Gabarra}, L. and {Mart{\'\i}n-Fleitas}, J. and {Risso}, I. and {Scottez}, V. and {Viel}, M.},
        title = "{Euclid: Photometric redshift calibration performance with the clustering-redshifts technique in the Flagship 2 simulation}",
      journal = {\aap},
         year = 2025,
        month = oct,
       volume = {702},
          eid = {A155},
        pages = {A155},
          doi = {10.1051/0004-6361/202555551},
archivePrefix = {arXiv},
       eprint = {2505.10416},
 primaryClass = {astro-ph.CO},
       adsurl = {https://ui.adsabs.harvard.edu/abs/2025A&A...702A.155D}
}

@ARTICLE{4MOST,
       author = {{Verdier}, Aur{\'e}lien and {Rocher}, Antoine and {Bandi}, Behnood and {Richard}, Johan and {Roukema}, Boudewijn F. and {Loveday}, Jon and {Tempel}, Elmo and {Bilicki}, Maciej and {Kneib}, Jean-Paul and {Guitton}, Mathilde},
        title = "{The 4MOST-Cosmology Redshift Survey: target selection of bright galaxies and luminous red galaxies}",
      journal = {\mnras},
         year = 2026,
        month = jan,
       volume = {545},
       number = {3},
          eid = {staf2116},
        pages = {staf2116},
          doi = {10.1093/mnras/staf2116},
archivePrefix = {arXiv},
       eprint = {2508.07311},
 primaryClass = {astro-ph.CO},
       adsurl = {https://ui.adsabs.harvard.edu/abs/2026MNRAS.545f2116V}
}

@ARTICLE{Grandis25,
       author = {{Grandis}, S. and {Costanzi}, M. and {Mohr}, J.~J. and {Bleem}, L.~E. and {Wu}, H.-Y. and {Aguena}, M. and {Allam}, S. and {Andrade-Oliveira}, F. and {Bocquet}, S. and {Brooks}, D. and {Carnero Rosell}, A. and {Carretero}, J. and {da Costa}, L.~N. and {Pereira}, M.~E.~S. and {Davis}, T.~M. and {Desai}, S. and {Diehl}, H.~T. and {Doel}, P. and {Everett}, S. and {Flaugher}, B. and {Frieman}, J. and {Garc{\'\i}a-Bellido}, J. and {Gaztanaga}, E. and {Gruen}, D. and {Gruendl}, R.~A. and {Gutierrez}, G. and {Hinton}, S.~R. and {Hlacacek-Larrondo}, J. and {Hollowood}, D.~L. and {Honscheid}, K. and {James}, D.~J. and {Klein}, M. and {Marshall}, J.~L. and {Mena-Fern{\'a}ndez}, J. and {Miquel}, R. and {Palmese}, A. and {Plazas Malag{\'o}n}, A.~A. and {Reichardt}, C.~L. and {Romer}, A.~K. and {Samuroff}, S. and {Sanchez Cid}, D. and {Sanchez}, E. and {Santiago}, B. and {Saro}, A. and {Sevilla-Noarbe}, I. and {Smith}, M. and {Soares-Santos}, M. and {Sommer}, M.~W. and {Suchyta}, E. and {Tarle}, G. and {To}, C. and {Tucker}, D.~L. and {Weaverdyck}, N. and {Weller}, J. and {Wiseman}, P.},
        title = "{Selection function of clusters in Dark Energy Survey year 3 data from cross-matching with South Pole Telescope detections}",
      journal = {\aap},
         year = 2025,
        month = aug,
       volume = {700},
          eid = {A15},
        pages = {A15},
          doi = {10.1051/0004-6361/202554177},
archivePrefix = {arXiv},
       eprint = {2502.12914},
 primaryClass = {astro-ph.CO},
       adsurl = {https://ui.adsabs.harvard.edu/abs/2025A&A...700A..15G}
}

@ARTICLE{Gonzalez-Perez20,
       author = {{Gonzalez-Perez}, V. and {Cui}, W. and {Contreras}, S. and {Baugh}, C.~M. and {Comparat}, J. and {Griffin}, A.~J. and {Helly}, J. and {Knebe}, A. and {Lacey}, C. and {Norberg}, P.},
        title = "{Do model emission line galaxies live in filaments at z {\ensuremath{\sim}} 1?}",
      journal = {\mnras},
         year = 2020,
        month = oct,
       volume = {498},
       number = {2},
        pages = {1852-1870},
          doi = {10.1093/mnras/staa2504},
archivePrefix = {arXiv},
       eprint = {2001.06560},
 primaryClass = {astro-ph.GA},
       adsurl = {https://ui.adsabs.harvard.edu/abs/2020MNRAS.498.1852G}
}

@ARTICLE{Gatti22WZ,
       author = {{Gatti}, M. and {Giannini}, G. and {Bernstein}, G.~M. and {Alarcon}, A. and {Myles}, J. and {Amon}, A. and {Cawthon}, R. and {Troxel}, M. and {DeRose}, J. and {Everett}, S. and {Ross}, A.~J. and {Rykoff}, E.~S. and {Elvin-Poole}, J. and {Cordero}, J. and {Harrison}, I. and {Sanchez}, C. and {Prat}, J. and {Gruen}, D. and {Lin}, H. and {Crocce}, M. and {Rozo}, E. and {Abbott}, T.~M.~C. and {Aguena}, M. and {Allam}, S. and {Annis}, J. and {Avila}, S. and {Bacon}, D. and {Bertin}, E. and {Brooks}, D. and {Burke}, D.~L. and {Rosell}, A. Carnero and {Kind}, M. Carrasco and {Carretero}, J. and {Castander}, F.~J. and {Choi}, A. and {Conselice}, C. and {Costanzi}, M. and {Crocce}, M. and {da Costa}, L.~N. and {Pereira}, M.~E.~S. and {Dawson}, K. and others and {DES Collaboration}},
        title = "{Dark Energy Survey Year 3 Results: clustering redshifts - calibration of the weak lensing source redshift distributions with redMaGiC and BOSS/eBOSS}",
      journal = {\mnras},
         year = 2022,
        month = feb,
       volume = {510},
       number = {1},
        pages = {1223-1247},
          doi = {10.1093/mnras/stab3311},
archivePrefix = {arXiv},
       eprint = {2012.08569},
 primaryClass = {astro-ph.CO},
       adsurl = {https://ui.adsabs.harvard.edu/abs/2022MNRAS.510.1223G}
}

@ARTICLE{Ivezic19,
       author = {{Ivezi{\'c}}, {\v{Z}}eljko and {Kahn}, Steven M. and {Tyson}, J. Anthony and {Abel}, Bob and {Acosta}, Emily and {Allsman}, Robyn and {Alonso}, David and {AlSayyad}, Yusra and {Anderson}, Scott F. and {Andrew}, John and {Angel}, James Roger P. and {Angeli}, George Z. and {Ansari}, Reza and {Antilogus}, Pierre and {Araujo}, Constanza and {Armstrong}, Robert and {Arndt}, Kirk T. and {Astier}, Pierre and {Aubourg}, {\'E}ric and {Auza}, Nicole and {Axelrod}, Tim S. and {Bard}, Deborah J. and {Barr}, Jeff D. and {Barrau}, Aurelian and {Bartlett}, James G. and {Bauer}, Amanda E. and others},
        title = "{LSST: From Science Drivers to Reference Design and Anticipated Data Products}",
      journal = {\apj},
         year = 2019,
        month = mar,
       volume = {873},
       number = {2},
          eid = {111},
        pages = {111},
          doi = {10.3847/1538-4357/ab042c},
archivePrefix = {arXiv},
       eprint = {0805.2366},
 primaryClass = {astro-ph},
       adsurl = {https://ui.adsabs.harvard.edu/abs/2019ApJ...873..111I}
}

@ARTICLE{Mellier25,
       author = {{Euclid Collaboration} and {Mellier}, Y. and {Abdurro'uf} and {Acevedo Barroso}, J.~A. and {Ach{\'u}carro}, A. and {Adamek}, J. and {Adam}, R. and {Addison}, G.~E. and {Aghanim}, N. and {Aguena}, M. and {Ajani}, V. and {Akrami}, Y. and {Al-Bahlawan}, A. and {Alavi}, A. and {Albuquerque}, I.~S. and {Alestas}, G. and {Alguero}, G. and {Allaoui}, A. and {Allen}, S.~W. and {Allevato}, V. and {Alonso-Tetilla}, A.~V. and {Altieri}, B. and {Alvarez-Candal}, A. and {Alvi}, S. and {Amara}, A. and {Amendola}, L. and {Amiaux}, J. and {Andika}, I.~T. and {Andreon}, S. and {Andrews}, A. and {Angora}, G. and {Angulo}, R.~E. and {Annibali}, F. and {Anselmi}, A. and {Anselmi}, S. and {Arcari}, S. and {Archidiacono}, M. and {Aric{\`o}}, G. and {Arnaud}, M. and {Arnouts}, S. and {Asgari}, M. and {Asorey}, J. and {Atayde}, L. and {Atek}, H. and {Atrio-Barandela}, F. and {Aubert}, M. and {Aubourg}, E. and {Auphan}, T. and {Auricchio}, N. and {Aussel}, B. and {Aussel}, H. and {Avelino}, P.~P. and {Avgoustidis}, A. and {Avila}, S. and {Awan}, S. and {Azzollini}, R. and others},
        title = "{Euclid: I. Overview of the Euclid mission}",
      journal = {\aap},
         year = 2025,
        month = may,
       volume = {697},
          eid = {A1},
        pages = {A1},
          doi = {10.1051/0004-6361/202450810},
archivePrefix = {arXiv},
       eprint = {2405.13491},
 primaryClass = {astro-ph.CO},
       adsurl = {https://ui.adsabs.harvard.edu/abs/2025A&A...697A...1E}
}

@ARTICLE{DESI,
       author = {{DESI Collaboration} and {Aghamousa}, Amir and {Aguilar}, Jessica and {Ahlen}, Steve and {Alam}, Shadab and {Allen}, Lori E. and {Allende Prieto}, Carlos and {Annis}, James and {Bailey}, Stephen and {Balland}, Christophe and {Ballester}, Otger and {Baltay}, Charles and {Beaufore}, Lucas and {Bebek}, Chris and {Beers}, Timothy C. and {Bell}, Eric F. and {Bernal}, Jos{\'e} Luis and {Besuner}, Robert and {Beutler}, Florian and {Blake}, Chris and {Bleuler}, Hannes and {Blomqvist}, Michael and {Blum}, Robert and {Bolton}, Adam S. and {Briceno}, Cesar and {Brooks}, David and {Brownstein}, Joel R. and {Buckley-Geer}, Elizabeth and {Burden}, Angela and {Burtin}, Etienne and {Busca}, Nicolas G. and others},
        title = "{The DESI Experiment Part I: Science,Targeting, and Survey Design}",
      journal = {arXiv e-prints},
         year = 2016,
        month = oct,
          eid = {arXiv:1611.00036},
        pages = {arXiv:1611.00036},
          doi = {10.48550/arXiv.1611.00036},
archivePrefix = {arXiv},
       eprint = {1611.00036},
 primaryClass = {astro-ph.IM},
       adsurl = {https://ui.adsabs.harvard.edu/abs/2016arXiv161100036D}
}

@ARTICLE{Raichoor23ELG,
       author = {{Raichoor}, A. and {Moustakas}, J. and {Newman}, Jeffrey A. and {Karim}, T. and {Ahlen}, S. and {Alam}, Shadab and {Bailey}, S. and {Brooks}, D. and {Dawson}, K. and {de la Macorra}, A. and {de Mattia}, A. and {Dey}, A. and {Dey}, Biprateep and {Dhungana}, G. and {Eftekharzadeh}, S. and {Eisenstein}, D.~J. and {Fanning}, K. and {Font-Ribera}, A. and {Garc{\'\i}a-Bellido}, J. and {Gazta{\~n}aga}, E. and {A Gontcho}, S. Gontcho and {Guy}, J. and {Honscheid}, K. and {Ishak}, M. and {Kehoe}, R. and {Kisner}, T. and {Kremin}, Anthony and {Lan}, Ting-Wen and {Landriau}, M. and {Le Guillou}, L. and {Levi}, Michael E. and {Magneville}, C. and {Manera}, M. and {Martini}, P. and {Meisner}, Aaron M. and {Myers}, Adam D. and {Nie}, Jundan and {Palanque-Delabrouille}, N. and {Percival}, W.~J. and {Poppett}, C. and {Prada}, F. and {Ross}, A.~J. and {Ruhlmann-Kleider}, V. and {Sabiu}, C.~G. and {Schlafly}, E.~F. and {Schlegel}, D. and {Tarl{\'e}}, Gregory and {Weaver}, B.~A. and {Y{\`e}che}, Christophe and {Zhou}, Rongpu and {Zhou}, Zhimin and {Zou}, H.},
        title = "{Target Selection and Validation of DESI Emission Line Galaxies}",
      journal = {\aj},
         year = 2023,
        month = mar,
       volume = {165},
       number = {3},
          eid = {126},
        pages = {126},
          doi = {10.3847/1538-3881/acb213},
archivePrefix = {arXiv},
       eprint = {2208.08513},
 primaryClass = {astro-ph.CO},
       adsurl = {https://ui.adsabs.harvard.edu/abs/2023AJ....165..126R}
}

@ARTICLE{Zhou23LRG,
       author = {{Zhou}, Rongpu and {Dey}, Biprateep and {Newman}, Jeffrey A. and {Eisenstein}, Daniel J. and {Dawson}, K. and {Bailey}, S. and {Berti}, A. and {Guy}, J. and {Lan}, Ting-Wen and {Zou}, H. and {Aguilar}, J. and {Ahlen}, S. and {Alam}, Shadab and {Brooks}, D. and {de la Macorra}, A. and {Dey}, A. and {Dhungana}, G. and {Fanning}, K. and {Font-Ribera}, A. and {Gontcho}, S. Gontcho A. and {Honscheid}, K. and {Ishak}, Mustapha and {Kisner}, T. and {Kov{\'a}cs}, A. and {Kremin}, A. and {Landriau}, M. and {Levi}, Michael E. and {Magneville}, C. and {Manera}, Marc and {Martini}, P. and {Meisner}, Aaron M. and {Miquel}, R. and {Moustakas}, J. and {Myers}, Adam D. and {Nie}, Jundan and {Palanque-Delabrouille}, N. and {Percival}, W.~J. and {Poppett}, C. and {Prada}, F. and {Raichoor}, A. and {Ross}, A.~J. and {Schlafly}, E. and {Schlegel}, D. and {Schubnell}, M. and {Tarl{\'e}}, Gregory and {Weaver}, B.~A. and {Wechsler}, R.~H. and {Y{\'e}che}, Christophe and {Zhou}, Zhimin},
        title = "{Target Selection and Validation of DESI Luminous Red Galaxies}",
      journal = {\aj},
         year = 2023,
        month = feb,
       volume = {165},
       number = {2},
          eid = {58},
        pages = {58},
          doi = {10.3847/1538-3881/aca5fb},
archivePrefix = {arXiv},
       eprint = {2208.08515},
 primaryClass = {astro-ph.CO},
       adsurl = {https://ui.adsabs.harvard.edu/abs/2023AJ....165...58Z}
}

@ARTICLE{Eisenstein01,
       author = {{Eisenstein}, Daniel J. and {Annis}, James and {Gunn}, James E. and {Szalay}, Alexander S. and {Connolly}, Andrew J. and {Nichol}, R.~C. and {Bahcall}, Neta A. and {Bernardi}, Mariangela and {Burles}, Scott and {Castander}, Francisco J. and {Fukugita}, Masataka and {Hogg}, David W. and {Ivezi{\'c}}, {\v{Z}}eljko and {Knapp}, G.~R. and {Lupton}, Robert H. and {Narayanan}, Vijay and {Postman}, Marc and {Reichart}, Daniel E. and {Richmond}, Michael and {Schneider}, Donald P. and {Schlegel}, David J. and {Strauss}, Michael A. and {SubbaRao}, Mark and {Tucker}, Douglas L. and {Vanden Berk}, Daniel and {Vogeley}, Michael S. and {Weinberg}, David H. and {Yanny}, Brian},
        title = "{Spectroscopic Target Selection for the Sloan Digital Sky Survey: The Luminous Red Galaxy Sample}",
      journal = {\aj},
         year = 2001,
        month = nov,
       volume = {122},
       number = {5},
        pages = {2267-2280},
          doi = {10.1086/323717},
archivePrefix = {arXiv},
       eprint = {astro-ph/0108153},
 primaryClass = {astro-ph},
       adsurl = {https://ui.adsabs.harvard.edu/abs/2001AJ....122.2267E}
}

@ARTICLE{Raichoor17,
       author = {{Raichoor}, A. and {Comparat}, J. and {Delubac}, T. and {Kneib}, J.-P. and {Y{\`e}che}, Ch and {Dawson}, K.~S. and {Percival}, W.~J. and {Dey}, A. and {Lang}, D. and {Schlegel}, D.~J. and {Gorgoni}, C. and {Bautista}, J. and {Brownstein}, J.~R. and {Mariappan}, V. and {Seo}, H.-J. and {Tinker}, J.~L. and {Ross}, A.~J. and {Wang}, Y. and {Zhao}, G.-B. and {Moustakas}, J. and {Palanque-Delabrouille}, N. and {Jullo}, E. and {Newmann}, J.~A. and {Prada}, F. and {Zhu}, G.~B.},
        title = "{The SDSS-IV extended Baryon Oscillation Spectroscopic Survey: final emission line galaxy target selection}",
      journal = {\mnras},
         year = 2017,
        month = nov,
       volume = {471},
       number = {4},
        pages = {3955-3973},
          doi = {10.1093/mnras/stx1790},
archivePrefix = {arXiv},
       eprint = {1704.00338},
 primaryClass = {astro-ph.CO},
       adsurl = {https://ui.adsabs.harvard.edu/abs/2017MNRAS.471.3955R}
}

@ARTICLE{ROTAC25,
       author = {{Zasowski}, Gail and others},
        title = "{Roman Observations Time Allocation Committee: Final Report and Recommendations}",
      journal = {arXiv e-prints},
         year = 2025,
        month = may,
          eid = {arXiv:2505.10574},
        pages = {arXiv:2505.10574},
          doi = {10.48550/arXiv.2505.10574},
archivePrefix = {arXiv},
       eprint = {2505.10574},
 primaryClass = {astro-ph.IM},
       adsurl = {https://ui.adsabs.harvard.edu/abs/2025arXiv250510574O}
}

@article{Nde26,
  title = {Impact of projection-induced optical selection bias on the weak lensing mass calibration of galaxy clusters},
  author = {Nde, Titus Nyarko and Wu, Hao-Yi and Cao, Shulei and Kamau, Gladys Muthoni and Tamosiunas, Andrius and To, Chun-Hao and Zhou, Conghao},
  journal = {Phys. Rev. D},
  volume = {113},
  issue = {6},
  pages = {063559},
  numpages = {8},
  year = {2026},
  month = {Mar},
  publisher = {American Physical Society},
  doi = {10.1103/h9xq-k9gh},
  url = {https://link.aps.org/doi/10.1103/h9xq-k9gh}
}

@ARTICLE{Cao25,
       author = {{Cao}, Shulei and {Wu}, Hao-Yi and {Costanzi}, Matteo and {Farahi}, Arya and {Grandis}, Sebastian and {Weinberg}, David H. and {Evrard}, August E. and {Rozo}, Eduardo and {Salcedo}, Andr{\'e}s N. and {To}, Chun-Hao and {Yang}, Lei and {Zhou}, Conghao and {DES Collaboration}},
        title = "{Association between optically identified galaxy clusters and the underlying dark matter halos}",
      journal = {\prd},
         year = 2025,
        month = aug,
       volume = {112},
       number = {4},
          eid = {043517},
        pages = {043517},
          doi = {10.1103/r7tt-bzs7},
archivePrefix = {arXiv},
       eprint = {2506.17526},
 primaryClass = {astro-ph.CO},
       adsurl = {https://ui.adsabs.harvard.edu/abs/2025PhRvD.112d3517C}
}

@ARTICLE{Liu22eFEDS,
       author = {{Liu}, A. and {Bulbul}, E. and {Ghirardini}, V. and {Liu}, T. and {Klein}, M. and {Clerc}, N. and {{\"O}zsoy}, Y. and {Ramos-Ceja}, M.~E. and {Pacaud}, F. and {Comparat}, J. and {Okabe}, N. and {Bahar}, Y.~E. and {Biffi}, V. and {Brunner}, H. and {Br{\"u}ggen}, M. and {Buchner}, J. and {Ider Chitham}, J. and {Chiu}, I. and {Dolag}, K. and {Gatuzz}, E. and {Gonzalez}, J. and {Hoang}, D.~N. and {Lamer}, G. and {Merloni}, A. and {Nandra}, K. and {Oguri}, M. and {Ota}, N. and {Predehl}, P. and {Reiprich}, T.~H. and {Salvato}, M. and {Schrabback}, T. and {Sanders}, J.~S. and {Seppi}, R. and {Thibaud}, Q.},
        title = "{The eROSITA Final Equatorial-Depth Survey (eFEDS). Catalog of galaxy clusters and groups}",
      journal = {\aap},
         year = 2022,
        month = may,
       volume = {661},
          eid = {A2},
        pages = {A2},
          doi = {10.1051/0004-6361/202141120},
archivePrefix = {arXiv},
       eprint = {2106.14518},
 primaryClass = {astro-ph.CO},
       adsurl = {https://ui.adsabs.harvard.edu/abs/2022A&A...661A...2L}
}

@ARTICLE{Bulbul24,
       author = {{Bulbul}, E. and {Liu}, A. and {Kluge}, M. and {Zhang}, X. and {Sanders}, J.~S. and {Bahar}, Y.~E. and {Ghirardini}, V. and {Artis}, E. and {Seppi}, R. and {Garrel}, C. and {Ramos-Ceja}, M.~E. and {Comparat}, J. and {Balzer}, F. and {B{\"o}ckmann}, K. and {Br{\"u}ggen}, M. and {Clerc}, N. and {Dennerl}, K. and {Dolag}, K. and {Freyberg}, M. and {Grandis}, S. and {Gruen}, D. and {Kleinebreil}, F. and {Krippendorf}, S. and {Lamer}, G. and {Merloni}, A. and {Migkas}, K. and {Nandra}, K. and {Pacaud}, F. and {Predehl}, P. and {Reiprich}, T.~H. and {Schrabback}, T. and {Veronica}, A. and {Weller}, J. and {Zelmer}, S.},
        title = "{The SRG/eROSITA All-Sky Survey. The first catalog of galaxy clusters and groups in the Western Galactic Hemisphere}",
      journal = {\aap},
         year = 2024,
        month = may,
       volume = {685},
          eid = {A106},
        pages = {A106},
          doi = {10.1051/0004-6361/202348264},
archivePrefix = {arXiv},
       eprint = {2402.08452},
 primaryClass = {astro-ph.CO},
       adsurl = {https://ui.adsabs.harvard.edu/abs/2024A&A...685A.106B}
}

@ARTICLE{DESY3CL3x2pt,
       author = {{Abbott}, T.~M.~C. and {Aguena}, M. and {Alarcon}, A. and {Amon}, A. and {Anbajagane}, D. and {Andrade-Oliveira}, F. and {Avila}, S. and {Allam}, S.~S. and {Bacon}, D. and {Becker}, M.~R. and {Bhargava}, S. and {Blazek}, J. and {Bocquet}, S. and {Brooks}, D. and others},
        title = "{Dark energy survey year 3 results: Cosmological constraints from cluster abundances, weak lensing, and galaxy clustering}",
      journal = {\prd},
         year = 2025,
        month = oct,
       volume = {112},
       number = {8},
          eid = {083535},
        pages = {083535},
          doi = {10.1103/3dzh-d8f5},
archivePrefix = {arXiv},
       eprint = {2503.13632},
 primaryClass = {astro-ph.CO},
       adsurl = {https://ui.adsabs.harvard.edu/abs/2025PhRvD.112h3535A}
}

@ARTICLE{Lee25,
       author = {{Lee}, Andy and {Wu}, Hao-Yi and {Salcedo}, Andr{\'e}s N. and {Sunayama}, Tomomi and {Costanzi}, Matteo and {Myles}, Justin and {Cao}, Shulei and {Rozo}, Eduardo and {To}, Chun-Hao and {Weinberg}, David H. and {Yang}, Lei and {Zhou}, Conghao},
        title = "{Optical galaxy cluster mock catalogs with realistic projection effects: Validations with the SDSS clusters}",
      journal = {\prd},
         year = 2025,
        month = mar,
       volume = {111},
       number = {6},
          eid = {063502},
        pages = {063502},
          doi = {10.1103/PhysRevD.111.063502},
archivePrefix = {arXiv},
       eprint = {2410.02497},
 primaryClass = {astro-ph.CO},
       adsurl = {https://ui.adsabs.harvard.edu/abs/2025PhRvD.111f3502L}
}

@ARTICLE{WenHan24,
       author = {{Wen}, Z.~L. and {Han}, J.~L.},
        title = "{A Catalog of 1.58 Million Clusters of Galaxies Identified from the DESI Legacy Imaging Surveys}",
      journal = {\apjs},
         year = 2024,
        month = jun,
       volume = {272},
       number = {2},
          eid = {39},
        pages = {39},
          doi = {10.3847/1538-4365/ad409d},
archivePrefix = {arXiv},
       eprint = {2404.02002},
 primaryClass = {astro-ph.CO},
       adsurl = {https://ui.adsabs.harvard.edu/abs/2024ApJS..272...39W}
}

@ARTICLE{Grishin23,
       author = {{Grishin}, Kirill and {Mei}, Simona and {Ili{\'c}}, St{\'e}phane},
        title = "{YOLO-CL: Galaxy cluster detection in the SDSS with deep machine learning}",
      journal = {\aap},
         year = 2023,
        month = sep,
       volume = {677},
          eid = {A101},
        pages = {A101},
          doi = {10.1051/0004-6361/202345976},
archivePrefix = {arXiv},
       eprint = {2301.09657},
 primaryClass = {astro-ph.CO},
       adsurl = {https://ui.adsabs.harvard.edu/abs/2023A&A...677A.101G}
}

@ARTICLE{Yantovski-Barth24,
       author = {{Yantovski-Barth}, M.~J. and {Newman}, Jeffrey A. and {Dey}, Biprateep and {Andrews}, Brett H. and {Eracleous}, Michael and {Golden-Marx}, Jesse and {Zhou}, Rongpu},
        title = "{The CluMPR galaxy cluster-finding algorithm and DESI legacy survey galaxy cluster catalogue}",
      journal = {\mnras},
         year = 2024,
        month = jun,
       volume = {531},
       number = {2},
        pages = {2285-2303},
          doi = {10.1093/mnras/stae956},
archivePrefix = {arXiv},
       eprint = {2307.10426},
 primaryClass = {astro-ph.CO},
       adsurl = {https://ui.adsabs.harvard.edu/abs/2024MNRAS.531.2285Y}
}

@ARTICLE{Cawthon22,
       author = {{Cawthon}, R. and {Elvin-Poole}, J. and {Porredon}, A. and {Crocce}, M. and {Giannini}, G. and {Gatti}, M. and {Ross}, A.~J. and {Rykoff}, E.~S. and {Carnero Rosell}, A. and {DeRose}, J. and {Lee}, S. and {Rodriguez-Monroy}, M. and {Amon}, A. and {Bechtol}, K. and {De Vicente}, J. and {Gruen}, D. and {Morgan}, R. and {Sanchez}, E. and {Sanchez}, J. and {Sevilla-Noarbe}, I. and {Abbott}, T.~M.~C. and {Aguena}, M. and {Allam}, S. and {Annis}, J. and {Avila}, S. and {Bacon}, D. and {Bertin}, E. and {Brooks}, D. and {Burke}, D.~L. and others and {DES Collaboration}},
        title = "{Dark Energy Survey Year 3 results: calibration of lens sample redshift distributions using clustering redshifts with BOSS/eBOSS}",
      journal = {\mnras},
         year = 2022,
        month = jul,
       volume = {513},
       number = {4},
        pages = {5517-5539},
          doi = {10.1093/mnras/stac1160},
archivePrefix = {arXiv},
       eprint = {2012.12826},
 primaryClass = {astro-ph.CO},
       adsurl = {https://ui.adsabs.harvard.edu/abs/2022MNRAS.513.5517C}
}

@ARTICLE{Capasso19,
       author = {{Capasso}, R. and {Mohr}, J.~J. and {Saro}, A. and {Biviano}, A. and {Clerc}, N. and {Finoguenov}, A. and {Grandis}, S. and {Collins}, C. and {Erfanianfar}, G. and {Damsted}, S. and {Kirkpatrick}, C. and {Kukkola}, A.},
        title = "{Mass calibration of the CODEX cluster sample using SPIDERS spectroscopy - I. The richness-mass relation}",
      journal = {\mnras},
         year = 2019,
        month = jun,
       volume = {486},
       number = {2},
        pages = {1594-1607},
          doi = {10.1093/mnras/stz931},
archivePrefix = {arXiv},
       eprint = {1812.06094},
 primaryClass = {astro-ph.CO},
       adsurl = {https://ui.adsabs.harvard.edu/abs/2019MNRAS.486.1594C}
}

@ARTICLE{Morrison17,
       author = {{Morrison}, C.~B. and {Hildebrandt}, H. and {Schmidt}, S.~J. and {Baldry}, I.~K. and {Bilicki}, M. and {Choi}, A. and {Erben}, T. and {Schneider}, P.},
        title = "{the-wizz: clustering redshift estimation for everyone}",
      journal = {\mnras},
         year = 2017,
        month = may,
       volume = {467},
       number = {3},
        pages = {3576-3589},
          doi = {10.1093/mnras/stx342},
archivePrefix = {arXiv},
       eprint = {1609.09085},
 primaryClass = {astro-ph.CO},
       adsurl = {https://ui.adsabs.harvard.edu/abs/2017MNRAS.467.3576M}
}

@ARTICLE{Choi16,
       author = {{Choi}, A. and {Heymans}, C. and {Blake}, C. and {Hildebrandt}, H. and {Duncan}, C.~A.~J. and {Erben}, T. and {Nakajima}, R. and {Van Waerbeke}, L. and {Viola}, M.},
        title = "{CFHTLenS and RCSLenS: testing photometric redshift distributions using angular cross-correlations with spectroscopic galaxy surveys}",
      journal = {\mnras},
         year = 2016,
        month = dec,
       volume = {463},
       number = {4},
        pages = {3737-3754},
          doi = {10.1093/mnras/stw2241},
archivePrefix = {arXiv},
       eprint = {1512.03626},
 primaryClass = {astro-ph.CO},
       adsurl = {https://ui.adsabs.harvard.edu/abs/2016MNRAS.463.3737C}
}

@ARTICLE{Menard13,
       author = {{M{\'e}nard}, Brice and {Scranton}, Ryan and {Schmidt}, Samuel and {Morrison}, Chris and {Jeong}, Donghui and {Budavari}, Tamas and {Rahman}, Mubdi},
        title = "{Clustering-based redshift estimation: method and application to data}",
      journal = {arXiv e-prints},
         year = 2013,
        month = mar,
          eid = {arXiv:1303.4722},
        pages = {arXiv:1303.4722},
          doi = {10.48550/arXiv.1303.4722},
archivePrefix = {arXiv},
       eprint = {1303.4722},
 primaryClass = {astro-ph.CO},
       adsurl = {https://ui.adsabs.harvard.edu/abs/2013arXiv1303.4722M}
}

@ARTICLE{McQuinnWhite13,
       author = {{McQuinn}, Matthew and {White}, Martin},
        title = "{On using angular cross-correlations to determine source redshift distributions}",
      journal = {\mnras},
         year = 2013,
        month = aug,
       volume = {433},
       number = {4},
        pages = {2857-2883},
          doi = {10.1093/mnras/stt914},
archivePrefix = {arXiv},
       eprint = {1302.0857},
 primaryClass = {astro-ph.CO},
       adsurl = {https://ui.adsabs.harvard.edu/abs/2013MNRAS.433.2857M}
}

@ARTICLE{Schmidt13,
       author = {{Schmidt}, Samuel J. and {M{\'e}nard}, Brice and {Scranton}, Ryan and {Morrison}, Christopher and {McBride}, Cameron K.},
        title = "{Recovering redshift distributions with cross-correlations: pushing the boundaries}",
      journal = {\mnras},
         year = 2013,
        month = jun,
       volume = {431},
       number = {4},
        pages = {3307-3318},
          doi = {10.1093/mnras/stt410},
archivePrefix = {arXiv},
       eprint = {1303.0292},
 primaryClass = {astro-ph.CO},
       adsurl = {https://ui.adsabs.harvard.edu/abs/2013MNRAS.431.3307S}
}

@ARTICLE{MatthewsNewman10,
       author = {{Matthews}, Daniel J. and {Newman}, Jeffrey A.},
        title = "{Reconstructing Redshift Distributions with Cross-correlations: Tests and an Optimized Recipe}",
      journal = {\apj},
         year = 2010,
        month = sep,
       volume = {721},
       number = {1},
        pages = {456-468},
          doi = {10.1088/0004-637X/721/1/456},
archivePrefix = {arXiv},
       eprint = {1003.0687},
 primaryClass = {astro-ph.CO},
       adsurl = {https://ui.adsabs.harvard.edu/abs/2010ApJ...721..456M}
}

@ARTICLE{Anbajagane25DECADE,
       author = {{Anbajagane}, D. and {Zhang}, Z. and {Chang}, C. and {Tan}, C.~Y. and {Adamow}, M. and {Secco}, L.~F. and {Becker}, M.~R. and {Ferguson}, P.~S. and {Drlica-Wagner}, A. and {Gruendl}, R.~A. and {Herron}, K. and {Tong}, A. and {Troxel}, M.~A. and {Sanchez-Cid}, D. and {Sevilla-Noarbe}, I. and {Chicoine}, N. and {Teixeira}, R. and {Alarcon}, A. and {Suson}, D. and {Alsina}, A.~N. and {Amon}, A. and {Bom}, C.~R. and {Carballo-Bello}, J.~A. and {Cerny}, W. and {Choi}, A. and {Choi}, Y. and {Doux}, C. and {Eckert}, K. and {Gatti}, M. and {Gruen}, D. and {Jarvis}, M. and {James}, D.~J. and {Kuropatkin}, N. and {Mart{\'\i}nez-V{\'a}zquez}, C.~E. and {Massana}, P. and {Mau}, S. and {McCullough}, J. and {Medina}, G.~E. and {Mutlu-Pakdil}, B. and {Navabi}, M. and {No{\"e}l}, N.~E.~D. and {Pace}, A.~B. and {Prat}, J. and {Raveri}, M. and {Riley}, A.~H. and {Rykoff}, E.~S. and {Sakowska}, J.~D. and {Sand}, D.~J. and {Santana-Silva}, L. and {Shin}, T. and {Soares-Santos}, M. and {Stringfellow}, G.~S. and {Vivas}, A.~K. and {Yamamoto}, M.},
        title = "{The DECADE cosmic shear project I: A new weak lensing shape catalog of 107 million galaxies}",
      journal = {The Open Journal of Astrophysics},
         year = 2025,
        month = oct,
       volume = {8},
        pages = {46158},
          doi = {10.33232/001c.146158},
archivePrefix = {arXiv},
       eprint = {2502.17674},
 primaryClass = {astro-ph.CO},
       adsurl = {https://ui.adsabs.harvard.edu/abs/2025OJAp....846158A}
}

@ARTICLE{Sunayama24,
       author = {{Sunayama}, Tomomi and {Miyatake}, Hironao and {Sugiyama}, Sunao and {More}, Surhud and {Li}, Xiangchong and {Dalal}, Roohi and {Rau}, Markus M. and {Shi}, Jingjing and {Chiu}, I. -non and {Shirasaki}, Masato and {Zhang}, Tianqing and {Nishizawa}, Atsushi J.},
        title = "{Optical cluster cosmology with SDSS redMaPPer clusters and HSC-Y3 lensing measurements}",
      journal = {\prd},
         year = 2024,
        month = oct,
       volume = {110},
       number = {8},
          eid = {083511},
        pages = {083511},
          doi = {10.1103/PhysRevD.110.083511},
archivePrefix = {arXiv},
       eprint = {2309.13025},
 primaryClass = {astro-ph.CO},
       adsurl = {https://ui.adsabs.harvard.edu/abs/2024PhRvD.110h3511S}
}

@ARTICLE{Bleem24,
       author = {{Bleem}, L.~E. and {Klein}, M. and {Abbot}, T.~M.~C. and {Ade}, P.~A.~R. and {Aguena}, M. and {Alves}, O. and {Anderson}, A.~J. and {Andrade-Oliveira}, F. and {Ansarinejad}, B. and {Archipley}, M. and {Ashby}, M.~L.~N. and {Austermann}, J.~E. and {Bacon}, D. and {Beall}, J.~A. and {Bender}, A.~N. and {Benson}, B.~A. and {Bianchini}, F. and {Bocquet}, S. and {Brooks}, D. and {Burke}, D.~L. and others},
        title = "{Galaxy Clusters Discovered via the Thermal Sunyaev-Zel'dovich Effect in the 500-square-degree SPTpol Survey}",
      journal = {The Open Journal of Astrophysics},
         year = 2024,
        month = feb,
       volume = {7},
          eid = {13},
        pages = {13},
          doi = {10.21105/astro.2311.07512},
archivePrefix = {arXiv},
       eprint = {2311.07512},
 primaryClass = {astro-ph.CO},
       adsurl = {https://ui.adsabs.harvard.edu/abs/2024OJAp....7E..13B}
}

@ARTICLE{To23Cardinal,
       author = {{To}, Chun-Hao and {DeRose}, Joseph and {Wechsler}, Risa H. and {Rykoff}, Eli and {Wu}, Hao-Yi and {Adhikari}, Susmita and {Krause}, Elisabeth and {Rozo}, Eduardo and {Weinberg}, David H.},
        title = "{Buzzard to Cardinal: Improved Mock Catalogs for Large Galaxy Surveys}",
      journal = {\apj},
         year = 2024,
        month = jan,
       volume = {961},
       number = {1},
          eid = {59},
        pages = {59},
          doi = {10.3847/1538-4357/ad0e61},
archivePrefix = {arXiv},
       eprint = {2303.12104},
 primaryClass = {astro-ph.CO},
       adsurl = {https://ui.adsabs.harvard.edu/abs/2024ApJ...961...59T}
}

@ARTICLE{vandenBusch20,
       author = {{van den Busch}, J.~L. and {Hildebrandt}, H. and {Wright}, A.~H. and {Morrison}, C.~B. and {Blake}, C. and {Joachimi}, B. and {Erben}, T. and {Heymans}, C. and {Kuijken}, K. and {Taylor}, E.~N.},
        title = "{Testing KiDS cross-correlation redshifts with simulations}",
      journal = {\aap},
         year = 2020,
        month = oct,
       volume = {642},
          eid = {A200},
        pages = {A200},
          doi = {10.1051/0004-6361/202038835},
archivePrefix = {arXiv},
       eprint = {2007.01846},
 primaryClass = {astro-ph.CO},
       adsurl = {https://ui.adsabs.harvard.edu/abs/2020A&A...642A.200V}
}

@ARTICLE{Johnson17,
       author = {{Johnson}, Andrew and {Blake}, Chris and {Amon}, Alexandra and {Erben}, Thomas and {Glazebrook}, Karl and {Harnois-Deraps}, Joachim and {Heymans}, Catherine and {Hildebrandt}, Hendrik and {Joudaki}, Shahab and {Klaes}, Dominik and {Kuijken}, Konrad and {Lidman}, Chris and {Marin}, Felipe A. and {McFarland}, John and {Morrison}, Christopher B. and {Parkinson}, David and {Poole}, Gregory B. and {Radovich}, Mario and {Wolf}, Christian},
        title = "{2dFLenS and KiDS: determining source redshift distributions with cross-correlations}",
      journal = {\mnras},
         year = 2017,
        month = mar,
       volume = {465},
       number = {4},
        pages = {4118-4132},
          doi = {10.1093/mnras/stw3033},
archivePrefix = {arXiv},
       eprint = {1611.07578},
 primaryClass = {astro-ph.CO},
       adsurl = {https://ui.adsabs.harvard.edu/abs/2017MNRAS.465.4118J}
}

@ARTICLE{Sunayama23,
       author = {{Sunayama}, Tomomi},
        title = "{Observational constraints of an anisotropic boost due to the projection effects using redMaPPer clusters}",
      journal = {\mnras},
         year = 2023,
        month = jun,
       volume = {521},
       number = {4},
        pages = {5064-5076},
          doi = {10.1093/mnras/stad786},
archivePrefix = {arXiv},
       eprint = {2205.03233},
 primaryClass = {astro-ph.CO},
       adsurl = {https://ui.adsabs.harvard.edu/abs/2023MNRAS.521.5064S}
}

@ARTICLE{Davis18,
       author = {{Davis}, C. and {Rozo}, E. and {Roodman}, A. and {Alarcon}, A. and {Cawthon}, R. and {Gatti}, M. and {Lin}, H. and {Miquel}, R. and {Rykoff}, E.~S. and {Troxel}, M.~A. and {Vielzeuf}, P. and {Abbott}, T.~M.~C. and {Abdalla}, F.~B. and {Allam}, S. and {Annis}, J. and {Bechtol}, K. and {Benoit-L{\'e}vy}, A. and {Bertin}, E. and {Brooks}, D. and {Buckley-Geer}, E. and {Burke}, D.~L. and others and {DES Collaboration}},
        title = "{Cross-correlation redshift calibration without spectroscopic calibration samples in DES Science Verification Data}",
      journal = {\mnras},
         year = 2018,
        month = jun,
       volume = {477},
       number = {2},
        pages = {2196-2208},
          doi = {10.1093/mnras/sty787},
archivePrefix = {arXiv},
       eprint = {1707.08256},
 primaryClass = {astro-ph.CO},
       adsurl = {https://ui.adsabs.harvard.edu/abs/2018MNRAS.477.2196D}
}

@ARTICLE{Wu22,
       author = {{Wu}, Hao-Yi and {Costanzi}, Matteo and {To}, Chun-Hao and {Salcedo}, Andr{\'e}s N. and {Weinberg}, David H. and {Annis}, James and {Bocquet}, Sebastian and {da Silva Pereira}, Maria Elidaiana and {DeRose}, Joseph and {Esteves}, Johnny and {Farahi}, Arya and {Grandis}, Sebastian and {Rozo}, Eduardo and {Rykoff}, Eli S. and {Varga}, Tam{\'a}s N. and {Wechsler}, Risa H. and {Zeng}, Chenxiao and {Zhang}, Yuanyuan and {Zhang}, Zhuowen and {DES Collaboration}},
        title = "{Optical selection bias and projection effects in stacked galaxy cluster weak lensing}",
      journal = {\mnras},
         year = 2022,
        month = sep,
       volume = {515},
       number = {3},
        pages = {4471-4486},
          doi = {10.1093/mnras/stac2048},
archivePrefix = {arXiv},
       eprint = {2203.05416},
 primaryClass = {astro-ph.CO},
       adsurl = {https://ui.adsabs.harvard.edu/abs/2022MNRAS.515.4471W}
}

@ARTICLE{Rines18,
       author = {{Rines}, Kenneth J. and {Geller}, Margaret J. and {Diaferio}, Antonaldo and {Hwang}, Ho Seong and {Sohn}, Jubee},
        title = "{HeCS-red: Dense Hectospec Surveys of redMaPPer-selected Clusters}",
      journal = {\apj},
         year = 2018,
        month = aug,
       volume = {862},
       number = {2},
          eid = {172},
        pages = {172},
          doi = {10.3847/1538-4357/aacd49},
archivePrefix = {arXiv},
       eprint = {1712.00212},
 primaryClass = {astro-ph.CO},
       adsurl = {https://ui.adsabs.harvard.edu/abs/2018ApJ...862..172R}
}

@ARTICLE{Wetzell22,
       author = {{Wetzell}, V. and {Jeltema}, T.~E. and {Hegland}, B. and {Everett}, S. and {Giles}, P.~A. and {Wilkinson}, R. and {Farahi}, A. and {Costanzi}, M. and {Hollowood}, D.~L. and {Upsdell}, E. and {Saro}, A. and {Myles}, J. and {Bermeo}, A. and {Bhargava}, S. and {Collins}, C.~A. and {Cross}, D. and others and {DES Collaboration}},
        title = "{Velocity dispersions of clusters in the Dark Energy Survey Y3 redMaPPer catalogue}",
      journal = {\mnras},
         year = 2022,
        month = aug,
       volume = {514},
       number = {4},
        pages = {4696-4717},
          doi = {10.1093/mnras/stac1623},
archivePrefix = {arXiv},
       eprint = {2107.07631},
 primaryClass = {astro-ph.CO},
       adsurl = {https://ui.adsabs.harvard.edu/abs/2022MNRAS.514.4696W}
}

@ARTICLE{Wechsler21,
       author = {{Wechsler}, Risa H. and {DeRose}, Joseph and {Busha}, Michael T. and {Becker}, Matthew R. and {Rykoff}, Eli and {Evrard}, August},
        title = "{ADDGALS: Simulated Sky Catalogs for Wide Field Galaxy Surveys}",
      journal = {\apj},
         year = 2022,
        month = jun,
       volume = {931},
       number = {2},
          eid = {145},
        pages = {145},
          doi = {10.3847/1538-4357/ac5b0a},
archivePrefix = {arXiv},
       eprint = {2105.12105},
 primaryClass = {astro-ph.CO},
       adsurl = {https://ui.adsabs.harvard.edu/abs/2022ApJ...931..145W}
}

@ARTICLE{LSSTSRD18,
       author = {{The LSST Dark Energy Science Collaboration} and {Mandelbaum}, Rachel and {Eifler}, Tim and {Hlo{\v{z}}ek}, Ren{\'e}e and {Collett}, Thomas and {Gawiser}, Eric and {Scolnic}, Daniel and {Alonso}, David and {Awan}, Humna and {Biswas}, Rahul and {Blazek}, Jonathan and {Burchat}, Patricia and {Chisari}, Nora Elisa and {Dell'Antonio}, Ian and {Digel}, Seth and others},
        title = "{The LSST Dark Energy Science Collaboration (DESC) Science Requirements Document}",
      journal = {arXiv e-prints},
         year = 2018,
        month = sep,
          eid = {arXiv:1809.01669},
        pages = {arXiv:1809.01669},
          doi = {10.48550/arXiv.1809.01669},
archivePrefix = {arXiv},
       eprint = {1809.01669},
 primaryClass = {astro-ph.CO},
       adsurl = {https://ui.adsabs.harvard.edu/abs/2018arXiv180901669T}
}

@ARTICLE{Myles25,
       author = {{Myles}, J. and {Gruen}, D. and {Jeltema}, T. and {Mantz}, A. and {Allen}, S. and {Fu}, S. and {Kremin}, A. and {Aguilar}, J. and {Ahlen}, S. and {Bianchi}, D. and {Brooks}, D. and {Castander}, F.~J. and {Claybaugh}, T. and {de la Macorra}, A. and {Dey}, Arjun and {Doel}, P. and {Ferraro}, S. and others},
        title = "{Spectroscopic Characterization of redMaPPer Galaxy Clusters with DESI}",
      journal = {\mnras},
         year = 2025,
        month = dec,
       volume = {544},
       number = {2},
        pages = {2080-2097},
          doi = {10.1093/mnras/staf1831},
archivePrefix = {arXiv},
       eprint = {2506.06249},
 primaryClass = {astro-ph.CO},
       adsurl = {https://ui.adsabs.harvard.edu/abs/2025MNRAS.544.2080M}
}

@article{Voit05,
	Adsurl = {http://adsabs.harvard.edu/abs/2005RvMP...77..207V},
	Author = {{Voit}, G.~M.},
	Doi = {10.1103/RevModPhys.77.207},
	Eprint = {arXiv:astro-ph/0410173},
	Journal = {Reviews of Modern Physics},
	Month = apr,
	Pages = {207-258},
	Title = {{Tracing cosmic evolution with clusters of galaxies}},
	Volume = 77,
	Year = 2005}
\end{document}